\documentstyle[11pt,aaspp4]{article}

\received{November 16, 1995} 
\accepted{April 30, 1996}

\slugcomment{To appear in the Astrophysical Journal} 

\newcommand{\bez}{\begin{eqnarray*}}
\newcommand{\eez}{\end{eqnarray*}}
\newcommand{\be}{\begin{equation}}
\newcommand{\ee}{\end{equation}}

\newcommand{\beq}{\begin{eqnarray}}
\newcommand{\eeq}{\end{eqnarray}}
\newcommand{\bc}{\begin{center}}
\newcommand{\ec}{\end{center}}

\newcommand{\ovl}{\overline}

\def\h{\hat}
\def\a218{\alpha_{2-18}}
\def\ep{\epsilon}
\def\ldiss{l_{\rm diss}}
\def\ls{l_{\rm s}}
\def\lc{l_{\rm c}} 
\def\ldisk{l_{\rm disk}} 
\def\lrepr{l_{\rm repr}}
\def\lrefl{l_{\rm refl}}
\def\Ldiss{L_{\rm diss}}
\def\Tdisk{T_{\rm disk}} 
\def\Ldisk{L_{\rm disk}} 
\def\Ls{L_{\rm s}}
\def\Lc{L_{\rm c}}
\def\Te{T_{\rm e}}
\def\me{m_{\rm e}}
\def\re{r_{\rm e}}
\def\ne{n_{\rm e}}
\def\CI{C^{\rm I}}
\def\CE{C^{\rm E}}
\def\Tbb{T_{\rm bb}}
\def\g{\gamma}

\def\rmd{{\rm d}}
\def\ergs{\;{\rm erg}}
\def\cm{\;{\rm cm}}

\def\sec{\;{\rm s}}
\def\sr{\;{\rm sr}}  
\def\taur{\tau_{\rm R}}
\def\taut{\tau_{\rm T}}
\def\taux{\tau_{x}} 
\def\sigmat{\sigma_{\rm T}}
\def\sigmacs{\sigma_{\rm cs}}
\def\refl{\rm refl}
\def\costh{\cos\theta}
\def\sinth{\sin\theta}

\newbox\grsign \setbox\grsign=\hbox{$>$} \newdimen\grdimen \grdimen=\ht\grsign
\newbox\simlessbox \newbox\simgreatbox \newbox\simpropbox
\setbox\simgreatbox=\hbox{\raise.5ex\hbox{$>$}\llap
     {\lower.5ex\hbox{$\sim$}}}\ht1=\grdimen\dp1=0pt
\setbox\simlessbox=\hbox{\raise.5ex\hbox{$<$}\llap
     {\lower.5ex\hbox{$\sim$}}}\ht2=\grdimen\dp2=0pt
\setbox\simpropbox=\hbox{\raise.5ex\hbox{$\propto$}\llap
     {\lower.5ex\hbox{$\sim$}}}\ht2=\grdimen\dp2=0pt
\def\simgreat{\mathrel{\copy\simgreatbox}}
\def\simless{\mathrel{\copy\simlessbox}}

\lefthead{Poutanen \& Svensson} 
\righthead{Two-Phase Pair Corona Model} 

\begin{document} 
 
\thispagestyle{empty}                

\title{The Two-Phase Pair Corona Model
for Active Galactic Nuclei and X-ray Binaries: 
How to Obtain Exact Solutions} 

\author{Juri Poutanen and  Roland Svensson}
\affil{Stockholm Observatory, S-133 36 Saltsj\"obaden, Sweden;  
 juri@astro.su.se, svensson@astro.su.se}

   
\begin{abstract} 
We consider two phase accretion disk-corona models for active galactic nuclei 
and some X-ray binaries.  We describe in detail how one can 
exactly solve the polarized radiative transfer and Comptonization using
the iterative  scattering method, while simultaneously 
solving the energy and pair balance equation for both the cold and hot phases. 
We take into account Compton scattering, photon-photon pair production, 
pair annihilation,  bremsstrahlung,   and double Compton scattering, 
as well as exact reflection from the cold disk.  We consider coronae having
slab geometry as well as coronae consisting of one or more well separated 
active regions of cylinder or  hemisphere geometry. 
 
The method is useful for determining the spectral intensity and the 
polarization emerging in different directions from disk-corona systems.
The code is tested against a Monte-Carlo code. We also compare with earlier,
less accurate, work. 
The method is more than an order of magnitude  faster than applying Monte 
Carlo methods to the same problem and has the potential of being used in 
spectral fitting software such as XSPEC. 
\end{abstract}


\section{Introduction}   

Both active galactic nuclei (AGN) and certain X-ray binaries (the galactic
black hole candidates, GBHC) show X-ray spectra extending into
the hard X-rays (e.g. Mushotzky, Done, \& Pounds 1993; Tanaka \& Levin 1995).
The X-ray spectra of Seyfert 1 galaxies show at least two components:
1) an intrinsic power law component with an intensity index, 
$\alpha \sim 0.9 - 1.0$, in the 2-18 keV range and a spectral cutoff
at a few hundred keV (Zdziarski et al. 1994; Madejski et al. 1995; 
Zdziarski et al. 1995), and 2) a superimposed reflection component
arising from reflection and reprocessing of the intrinsic power
law by cold opaque matter subtending  $\sim$ 1-2 $\pi$ solid angle
as viewed from the X-ray source (e.g. Nandra \& Pounds 1994).
GBHC such as Cyg X-1 and 1E1740.7-2942
show power law spectra extending 
up to hundreds of keV (e.g. Gilfanov et al. 1994; Tanaka \& Levin 1995). 
The characteristic features of reflection have been seen in GBHC 
as well (e.g. Done et al. 1992). 

AGN and GBHC are believed to be powered by accretion through 
an accretion disk. In the
unified model for AGN (e.g. Antonucci 1993), it is believed 
that we are viewing the disks in Seyfert 1 galaxies
more or less face on, while for Seyfert 2 galaxies we are viewing
the disk more or less edge on through a molecular torus.
In GBHC sources we are viewing the binary system along some given 
(possibly time dependent) direction.
The X-ray spectra indicate the existence of both hot X-ray emitting
and cold reflecting gas components. The exact geometry is not known,
but a currently popular model is the two-phase
disk-corona model (e.g. Haardt \& Maraschi 1991, 1993, hereafter HM93).
The black body disk radiation from the cold disk
 (in the EUV for AGN, and in the soft X-rays for GBHC)
enters the hot corona from one side and gets Comptonized into the X-rays.
Part of this X-ray radiation is incident on the cold disk and
is partly reflected but mainly reprocessed into soft black body radiation.
The remaining part forms the X-ray spectrum
leaving the disk-corona system. Both the black body
and the Comptonized spectrum are anisotropic so  observers at 
different viewing angles see different spectra (HM93).
It immediately clear that such models are neither homogeneous nor 
spherically symmetric. 
In order to correctly interpret observed X-ray spectra of AGN
and GBHC one needs to know
the theoretical spectra for mildly relativistic temperature and
for different viewing angles. 

Theoretical Comptonized spectra have been computed for two decades now.
Almost all work make simplifying assumptions that render them
useless for interpreting X-ray spectra from sources were anisotropic
effects are important. We briefly discuss standard methods for 
modeling Comptonized spectra from AGN and GBHC at mildly relativistic
temperatures.

One approach is to treat the photon and pair producing processes as
well as the energy and pair balance in great detail, but to
make large simplifications regarding the radiative transfer
using various prescriptions for the spectral shape
and using simple escape probabilities to get the photon density   
(e.g. Zdziarski 1985; Pietrini \& Krolik 1995). 
Such calculations can only give very approximate
relations between the typical spectral shape and  other parameters.

Other approaches are to do detailed radiative transfer using 
Monte Carlo methods for geometries such as slabs or spheres
(e.g. Hua \& Titarchuk 1995)
or to improve the analytical theory of Comptonization (e.g. Titarchuk 1994).
Normally the following simplifications are made:
1) processes other than Comptonization are neglected,
2) pair balance is not imposed,
3) angle dependence of output spectra is not considered, 
4) the soft photon injection is homogeneous throughout the source.
There are some exceptions. For example, Skibo et al. (1995) includes
bremsstrahlung, pair production, and pair balance, and Zdziarski et al.
(1994) assume the soft photons to be injected at one of the slab surfaces.
Most work neglect reflection by cold matter. The few papers
considering polarized radiative transfer 
(e.g. Sunyaev \& Titarchuk 1985; Haardt \& Matt 1993) 
use the Rayleigh matrix, which
is not valid for temperatures and photon energies above 50 keV.

Pioneering steps were taken by Haardt \& Maraschi (1991, 1993)
and Haardt (1993, 1994) to solve for angle-dependent spectra 
from disk-corona systems including reflection 
as well as energy and pair balance. The method treated
the first scattering order accurately, but assumed  higher
order scatterings to be isotropic. 
Compton recoil was neglected and therefore the
spectral cutoff at photon energies around and above $k\Te$
could not be treated. Furthermore,
the pair balance treatment was approximate as they adopted 
the semi-analytical theory 
of Zdziarski (1985) using prescribed spectra. 
Finally, when determining the reflected spectrum they
assumed the X-ray intensity incident on the
cold matter to be isotropic.
Further steps were taken by Stern et al. (1995a, b) who
used nonlinear Monte Carlo techniques to treat disk-corona systems
with different inhomogeneous coronal geometries. Here, exact
pair balance was imposed throughout the coronal region.
Furthermore, output spectra (including the cutoff)
had sufficiently good statistics to allow the dependence on the 
viewing angle to be studied.
These work found that anisotropic effects are very important
for spectra from disk-corona systems.
At mildly relativistic temperatures the first order scattering
of the soft disk radiation is suppressed in the face-on direction.
This causes the face-on spectrum to be harder up to the spectral peak
of the second order scattering, where an {\it anisotropy break}
appears. Above the break the Comptonized spectrum in all directions 
resembles the angle-averaged one. As shown in Stern et al. (1995b),
the anisotropy break easily appears in the 2-18 keV range
making $\a218$ strongly dependent on viewing angle.
The reflection component is also strongly anisotropic (e.g. Matt 1993; 
Magdziarz \& Zdziarski 1995; Poutanen, Nagendra, \& Svensson 1996). 

The disadvantages with the nonlinear Monte Carlo method is
that it is computer intensive requiring more than one hour per run on
a Sparc 20. In order to be able to do spectral fitting of
observed spectra one needs a much faster method to compute
accurate angle dependent spectra from disk-corona systems.
In the present paper we describe such a fast code based on the
iterative scattering method, i.e. where the radiative transfer
equation is solved for each scattering order (e.g. Sunyaev \& Titarchuk
1985). Most important radiation processes and photon-photon pair production
are included. Energy and pair balance can be imposed.
The reflection is accurately treated accounting for the
full angular dependence of the incident spectrum
(Poutanen et al. 1996). Both slab, cylinder, and hemisphere
geometries of the corona can be treated. Finally, the radiative transfer
is polarized, both as regards the Comptonization and the reflection.
The code has already been used to interpret the statistics of
observed X-ray spectral indices and compactnesses from
Seyfert 1 galaxies (Stern et al. 1995b).
The purpose of the present paper is to fully document
the methods used in the code in a selfcontained way.

In the remainder of the paper, we first describe the setup of the 
two-phase disk-corona model in \S~\ref{sec:setup}.
The methods of solving the radiative transfer equation in different 
geometries are considered in \S~\ref{sec:rte}. 
The energy and pair balance   and details 
of the iteration procedure are considered in \S~\ref{sec:baleq}. 
We compare our results with those of other available codes in 
\S~\ref{sec:compar}, where we also consider the accuracy and efficiency
of various approximations that can be employed in order to decrease
the computing time. 
Finally, we summarize our work in \S~\ref{sec:concl}.  
Expressions for the reaction rates and redistribution functions
(i.e. the Compton redistribution matrix) are given
in the Appendices.

\section{Setup}  \label{sec:setup}

We consider the simple two-phase disk-corona model, where a hot  
corona is located above an optically thick plane-parallel cold slab (``disk")
(e.g. Haardt \& Maraschi 1991, 1993; Haardt, Maraschi, \&
Ghisellini 1994).
The hot corona is either a plane-parallel slab (with vertical thickness $H$), 
or an 
active region   with the shape of a hemisphere (with radius $R$), or a cylinder
(with vertical height, $H$, and horizontal radius, $R$). 
We allow energy dissipation in both the corona and the cold disk. 
The radiation escaping from the cold disk consists of a soft component, 
and a reflected component. The soft flux is  equal to the sum
of the absorbed incident flux from the corona, and the flux due to local energy 
dissipation in the  cold disk. 
The spectral shape of these soft components 
is assumed to be Planckian with temperatures $\Tbb$ and $\Tdisk$, respectively
(note that $\Tbb>\Tdisk$). 
The shape of the reflected component is determined by the shape of 
incident coronal X-ray radiation 
(mainly resulting from Comptonization of the soft disk radiation) and 
the effects of photoelectric absorption and Compton scattering
in the cold disk (see, e.g., White, Lightman, \&  Zdziarski 1988; 
Magdziarz \& Zdziarski 1995).

We can treat coronae both with or without pairs. In this paper, we  mostly
consider the case of a pure pair corona without any background plasma.
The pure pair corona is  a consequence of photon-photon pair production 
above the cold disk. The electrons and positrons are assumed to
have a relativistic Maxwellian   distribution of the same temperature, 
$\Theta=k\Te/\me c^2$. The corona is assumed to be uniform
in temperature and pair density, and pair escape is neglected. 
If all power is dissipated in the corona then, for a given geometry, 
the two parameters: 
the total power dissipated in corona, $\Ldiss$, 
and the temperature, $\Tbb$, of the reprocessed radiation, 
uniquely determine the optical depth, $\taut$, and 
the coronal temperature, $\Theta$. If the radiation  produced internally
in the disk is not negligible, then two more parameters are important: 
the disk  temperature, $\Tdisk$, and 
the   ratio, $d=\Ldisk/\Ldiss$, where $\Ldisk$ is the luminosity 
that is produced internally in the disk and that enters the corona. 
For all geometries, $\taut$ is defined as the total vertical Thomson 
optical depth of the corona (along the symmetry axis
in the case of hemisphere geometry).
 
To solve the pair balance equation, the energy balance equations for 
the cold and hot phases,  and the radiative transfer  
in the corona self-consistently, we make use of an iteration procedure. 
To reduce computing time we choose to fix $\Theta$, which allow us to compute 
the thermal Compton redistribution matrix and cross section, and 
the coronal emissivities for pair annihilation and bremsstrahlung 
before doing the iterations. We 
then adjust $\taut$ and $\Ldiss$  until the radiation spectra from solving the 
radiative transfer  
satisfy the energy balance equations and the pair balance.

When solving the radiative transfer/Comptonization problem,
we account for the angular anisotropy of the radiation as well as its 
polarization properties. The reprocessing in the cold disk is described by 
Green's matrix (consisting of four Green's functions) for reflection, 
where we fully account for the Compton 
effect, photoelectric absorption,  and iron fluorescence, 
as well as for the angular and polarization
properties of the radiation incident on the cold disk (Poutanen et al. 1996). 
We reduce the time needed to compute the
reflection spectrum substantially using the precalculated Green's matrix 
and achieve better accuracy than all 
previous treatments of the problem (e.g.,  White  et al. 1988; 
Magdziarz \& Zdziarski 1995).

As additional photon sources and  cooling processes of the corona, we consider
electron-electron, positron-positron, and electron-positron bremsstrahlung, 
double Compton scattering, and pair annihilation.
We also account for photon absorption due to
pair production, which can be important in determining the spectral shapes at 
pair producing energies ($h\nu > \me c^2$) and, thus, in influencing 
the pair balance. 

The  radiative transfer equation is solved by expanding the radiation
field in scattering orders (the iterative scattering method, e.g.
Sunyaev \& Titarchuk 1985).
The intensity in a slab-corona is a function
of vertical position (i.e. the Thomson optical depth variable,
$\tau$), zenith angle and frequency but 
is azimuth-independent. In an active region, the intensity, of course, depends 
on the distance from the symmetry axis and the azimuth angle making the spatial 
part of the problem two-dimensional.  However, by averaging the radiation
field over horizontal layers in the active region, we convert the 2D-problem 
into a 1D-problem suitable for our 1D-code.
We discuss the accuracy of this conversion  in 
\S~\ref{sec:compar}. 

In the pair balance, we use a volume-averaged pair production rate.
In the energy balance equations, we need the total luminosities emerging
from the cold and the hot phases. 
Therefore we compute and sum the radiative fluxes emerging 
from all surfaces of the disk and the corona accounting for all radiative
transfer effects.

\section{Radiative Transfer} \label{sec:rte}
  
\subsection{Radiative Transfer in a Slab Corona}

Due to azimuthal symmetry and the absence  of sources of circular polarization,
the radiation field and the degree of polarization  at vertical position $z$ 
can be fully described by a Stokes vector consisting of two Stokes parameters 
(Chandrasekhar 1960) $\tilde{I}=\tilde{I}(z,x,\mu)
=(I,Q)^{\rm T}$, where $^{\rm T}$ denotes the transposed vector. 
The  radiative transfer equation describing the  propagation of polarized light
through a plane-parallel electron (and positron) atmosphere in steady-state 
can be written in the following form:  
\be \label{eq:radtr}
 \mu\frac{\rmd\tilde{I}(z,x,\mu)}{\rmd z} =
- (\ne\sigmacs(x)+\alpha_{\g\g}(z,x,\mu))\tilde{I}(z,x,\mu) +
\ne\sigmat \tilde{S}(z,x,\mu)  + \epsilon(x) , 
\ee
where $x\equiv h\nu/\me c^2$ is  the  photon energy,  
 $n_{\rm e}=n_++n_-$ is the total electron and positron density,
 $\mu$ is the cosine angle between the slab 
normal and the direction of photon propagation,
 $\sigmacs(x)\cm^{2}$ is the thermal Compton scattering cross section, 
 $\sigmat\cm^{2}$ is the Thomson cross section, 
$\alpha_{\g\g}(z,x,\mu)\cm^{-1}$ is the absorption coefficient due to 
photon-photon pair production, and $\tilde{S}(z,x,\mu)$ is  the 
 electron scattering source function.  The emissivity,
$\ep(x)=\ep_{++}+\ep_{--}+\ep_{+-}+\ep_{\rm ann}+\ep_{\rm DC}
\ergs \cm^{-1}  \sec^{-1} \sr^{-1}$, which is assumed to
be isotropic and homogeneous, includes all photon sources in 
the atmosphere, in our case electron-electron, positron-positron,
electron-positron bremsstrahlung, annihilation, and double Compton radiation.  
The expressions for the emissivities 
and absorption coefficients are given in the Appendices.  

 Using the following notation for the dimensionless 
intensity, source function, emissivities, scattering cross section 
and absorption coefficient: 
\bez 
I'=I\frac{H\sigmat}{\me c^3}, \quad S'=S\frac{H\sigmat}{\me c^3}, \quad 
\ep'=\ep\frac{H}{\ne\me c^3}, \quad\sigmacs'=\frac{\sigmacs}{\sigmat},  \quad
\alpha_{\g\g}'=\frac{\alpha_{\g\g}}{\ne\sigmat},   
\eez
and removing the primes, the radiative transfer equation 
can be written in the following dimensionless form: 
\be \label{eq:radtrdim}
 \mu\frac{\rmd\tilde{I}(\tau,x,\mu)}{\rmd \tau} =
- (\sigmacs(x)+\alpha_{\g\g}(\tau,x,\mu))\tilde{I}(\tau,x,\mu) +
 \tilde{S}(\tau,x,\mu)  + \ep(x) , 
\ee
where $\rmd\tau=\sigmat \ne \rmd z$ is the differential Thomson
optical depth. Hereafter we will only use dimensionless quantities 
(except in the Appendices). 
 
The Thomson optical depth of the slab is $\taut=H\sigmat \ne$. 
The boundary conditions at the upper  and 
lower surface of the slab are: 
\beq
\tilde{I}(\tau=\taut,x,-\mu)&=&0, \quad \mu>0,  \nonumber \\  
\tilde{I}(\tau=0,x,\mu)&=&\tilde{I}_{\rm in}(x,\mu) , 
\eeq
i.e. there is no  radiation incident at the upper surface, and 
the radiation incident 
at the lower surface consists of a reflected component, a soft reprocessed 
component,  and  a soft component internally produced in the disk: 
\be \label{eq:iinsl}
\tilde{I}_{\rm in}(x,\mu)=\tilde{I}_{\refl}(x,\mu)+c_{\rm bb}B_x(\Tbb)+
c_{\rm disk}B_x(\Tdisk)  .
\ee
The  reflected radiation, $\tilde{I}_{\refl}(x,\mu)$, from the cold disk can 
be  found as a  convolution of a reflection matrix 
(Green's matrix) $\h{G}(x,\mu;x_1,\mu_1)$   
with the incident radiation: 
\be \label{eq:reflec}
\tilde{I}_{\refl}(x,\mu)=\int_x^\infty \rmd x_1 \int_0^1 \rmd \mu_1 
\h{G}(x,\mu;x_1,\mu_1)\tilde{I}(\tau=0,x_1,-\mu_1) . 
\ee
The Green's matrix maps incoming radiation at ($x_1$, $-\mu_1$) into reflected
radiation at  ($x$, $\mu$). To compute Green's matrix we use the 
 method  developed by Poutanen et al. (1996).
The normalization  constants in front of the Planckian functions
in equation (\ref{eq:iinsl}) are determined by normalizing the black-body 
flux to the soft compactnesses, $\lrepr$ and $\ldisk$ (to be 
defined in \S~\ref{sec:energy}),  as:  
\be \label{eq:bbnorm}
 c_{\rm bb}\pi \int_0^\infty \rmd x B_x(\Tbb) =\lrepr ;  \qquad 
 c_{\rm disk}\pi \int_0^\infty \rmd x B_x(\Tdisk) =\ldisk . 
\ee

The thermal electron scattering source function, $\tilde{S}(\tau,x,\mu)$, 
 can be expressed in terms of the 
azimuth-averaged Compton redistribution matrix
 (see e.g. Poutanen \& Vilhu 1993): 
\be \label{eq:sours}
\tilde{S}(\tau,x,\mu)=x^2\int_0^{\infty}
 \frac{\rmd x_1}{x_1^2}\int_{-1}^{1} \rmd\mu_1 
\left( \begin{array}{cc}  
 {R}_{11}  &  {R}_{12} \\
 {R}_{21}  &  {R}_{22} 
\end{array} \right)   
 \tilde{I}(\tau,x_1,\mu_1) \, ,
\ee
or in operator form as
\be \label{eq:SRI}
\tilde{S} =  {\bf R} \tilde{I}  \, .  
\ee
The factor $x^2 / x_1^2$ in equation~(\ref{eq:sours})   appears because
we use the photon intensity instead of the photon occupation number 
to describe the radiation field  (see Nagirner \& Poutanen 1994).  
Expressions for the redistribution functions $R$ are given in 
Appendix~\ref{sec:sourcefunc}.

We  solve the integro-differential equation~(\ref{eq:radtrdim}) 
by expanding the
Stokes vector $\tilde{I}$ in scattering orders (Neumann series):
\be 
\tilde{I}=\sum_{k=0}^{\infty} \tilde{I}_k \, ,   
\ee 
where $\tilde{I}_k$ is the Stokes vector for photons having undergone $k$
scatterings (see, e.g., Sunyaev \& Titarchuk 1985). 
This expansion  converges  quickly for sufficiently 
small optical depths ($\taut\simless 1$).  
The source function for the non-scattered component consist of the flux 
incident on the corona at the bottom surface, $\tau = 0$,
and of the internal coronal sources:
\be
\tilde{S}_{0}(\tau,x,\mu)  =\mu 
\tilde{I}_{\rm in}(x,\mu) H(\mu) \delta(\tau) +   \ep(x) 
\ee
where $H(\mu)$ is the Heaviside function. 
The Stokes vectors, $\tilde{I}^\pm_k(\tau,x,\mu)=\tilde{I}_k(\tau,x,\pm\mu)$, 
for the upward and downward radiation and 
for all  scattering orders $k\ge 0$
are calculated employing the iteration formulae:
\beq \label{eq:iter1}
\tilde{I}^+_{k}(\tau,x,\mu) &=& 
\displaystyle \int_0^{\tau} \frac{\rmd \tau'}{\mu} \tilde{S}_{k}(\tau',x,\mu)
\exp\left\{-  
\int_{\tau'}^{\tau}\sigma(\tau'',x,\mu)
\frac{\rmd \tau''}{\mu}\right\}   
\, ,  \\ \label{eq:iter2}
\tilde{I}^-_{k}(\tau,x,\mu) &=& 
\displaystyle \int_{\tau}^{\taut} \frac{\rmd \tau'}{\mu} 
\tilde{S}_{k}(\tau',x,-\mu) \exp\left\{-\int_{\tau}^{\tau'}
\sigma(\tau'',x,-\mu) \frac{\rmd \tau''}{\mu}
 \right\}   \, , 
\eeq
where $\sigma(\tau,x,\mu)=\sigmacs(x)+\alpha_{\g\g}(\tau,x,\mu)$. 
Using equation~(\ref{eq:SRI})  the source function can be written as:
\be \label{eq:iter3}
\tilde{S}_{k+1} = {\bf R} \tilde{I}_{k} \,  .
\ee
This procedure gives the dependence of the Stokes parameters on frequency,
angle and optical depth. 
 Iterative methods where the calculations of
the spectral structure and of the angular polarization structure of 
the radiation field were separated 
(Sunyaev and Titarchuk 1985; Phillips and M\'esz\'aros 1986)
fail to obtain the frequency dependence of the Stokes
vectors for a given scattering order.
Substituting $\tau=\taut$ into equation~(\ref{eq:iter1}), and $\tau=0$ into 
equation~(\ref{eq:iter2}), we obtain the emergent Stokes vectors.

\subsection{Radiative Transfer in Cylinders}

In order to treat radiative transfer in cylindrical geometry we divide 
the cylinder into  horizontal spatial layers and average the  
computed radiation field over each layer (over the radius and the azimuthal 
directions) leaving only the dependence on the zenith angle. 
  To simplify the calculations we assume the soft (reprocessed and internally
produced in the cold disk) and reflected radiation to enter uniformly  at 
the base of the   cylinder.  Thus, we effectively convert the 2D-problem into a 
1D-problem. 

  The boundary conditions are the same as in the slab case, but 
the radiation incident on the base of the cylinder is now: 
\be \label{eq:cylbou}
\tilde{I}_{\rm in}(x,\mu)=g\tilde{I}_{\refl}(x,\mu)+c_{\rm bb}B_x(\Tbb)+
c_{\rm disk}B_x(\Tdisk) .
\ee
 The parameter $g$ is the fraction of the reprocessed and
reflected  radiation from the cold disk that enters the active region. 
In the case of slab geometry, $g=1$. For cylinders atop a cold disk $g\approx 
0.6$ if the vertical $\taut$  equals   the radial $\taur=R\ne\sigmat $, 
while $g \approx$ 0.45 for $\taut=2\taur$. The parameter Ê$\; g$
 is smaller for active regions detached from the cold disk. 
The normalization constants, $c_{\rm bb}$ and $c_{\rm disk}$,  are given by: 
\be \label{eq:bbnormcy}
 c_{\rm bb}\pi \int_0^\infty \rmd x B_x(\Tbb) = g\lrepr/\pi ; 
\qquad  c_{\rm disk}\pi \int_0^\infty \rmd x B_x(\Tdisk) = \ldisk/\pi .  
\ee
The expressions connecting the radiation field inside the cylinder with the 
source function are analogous to equations~(\ref{eq:iter1}) and (\ref{eq:iter2}): 
\beq \label{eq:cylsour1}
\tilde{I}^+_{k}(\tau,x,\mu) &= &
\displaystyle \int_0^{\tau} \frac{\rmd \tau'}{\mu} \tilde{S}_{k}(\tau',x,\mu)
\exp\left\{-  
\int_{\tau'}^{\tau}\sigma(\tau'',x,\mu)
\frac{\rmd \tau''}{\mu}\right\}   
\CI_+(\tau,\tau',\mu) \,  ,   \\\label{eq:cylsour2}
\tilde{I}^-_{k}(\tau,x,\mu) &= &
\displaystyle \int_{\tau}^{\taut} \frac{\rmd \tau'}{\mu}
 \tilde{S}_{k}(\tau',x,-\mu) \exp\left\{-  
\int_{\tau}^{\tau'}\sigma(\tau'',x,-\mu)
\frac{\rmd \tau''}{\mu}\right\}   \CI_-(\tau,\tau',\mu)
\, , 
\eeq 
where the correction factors, $\CI_\pm$, reduce the contributions from the 
source function at $\tau'$ to the radiation field at $\tau$ as compared to  
the slab case (note that $\mu>0$):  
\beq
\CI_\pm(\tau,\tau',\mu)&=&\left\{ \begin{array}{ll} 
0, & \mbox{if} \quad t\geq 1, \\
\frac{2}{\pi} \left( \arccos  t -t \sqrt{1-t^2} \right) , & \mbox{if} 
\quad t\leq 1, \end{array} \right.   \\ \label{eq:t}
t&=&\sqrt{1-\mu^2} |\tau-\tau'| / (\mu 2\taur)  .
\eeq
The equation for the source function (\ref{eq:iter3}) remains unchanged. 

The emerging (polarized) flux consists of two parts: 
  first, the radiation emerging through the top and the bottom of the cylinder,
and, second, the radiation emerging through the vertical surface.
The first part can trivially be found  from equation~(\ref{eq:cylsour1}) by 
multiplying the emerging intensity with $\mu\pi$ ($\pi$ appears because of the
definition of the compactness). 
The  second part is given by: 
\beq \label{eq:cylflux}
\tilde{F}^{\rm side,+}_k(x,\mu)&= &\frac{1}{\taur}\int_0^{\taut} \rmd \tau  
\int_0^{\tau} \frac{\rmd \tau'}{\mu} \tilde{S}_{k}(\tau',x,\mu)
\exp\left\{-  \int_{\tau'}^{\tau}\sigma(\tau'',x,\mu)
\frac{\rmd \tau''}{\mu}\right\}  \CE_+(\tau,\tau',\mu)
\,  ,    \\  \label{eq:cylfl2}
\tilde{F}^{\rm side,-}_k(x,\mu)&= &\frac{1}{\taur}\int_0^{\taut} \rmd \tau  
\int_{\tau}^{\taut} 
\frac{\rmd \tau'}{\mu} \tilde{S}_{k}(\tau',x,-\mu)
\exp\left\{-  \int_{\tau}^{\tau'}\sigma(\tau'',x,-\mu)
\frac{\rmd \tau''}{\mu}\right\} \CE_-(\tau,\tau',\mu) 
\, ,  
\eeq
where 
\be 
\CE_\pm(\tau,\tau',\mu)=
2\sqrt{1-\mu^2}  \sqrt{1-t^2}, \quad \mbox{if} \quad t<1 ,
\ee 
and equal to zero otherwise, $t$ is given by equation~(\ref{eq:t}).

\subsection{Radiative  Transfer in Hemispheres}

In the case of hemisphere geometry, we average  over 
the horizontal layers just as we did for cylinder geometry.   The incident 
radiation is given by equations (\ref{eq:cylbou}) and (\ref{eq:bbnormcy}), 
where the parameter $g \approx 0.7$ for hemispheres atop the cold disk.
The expressions (\ref{eq:cylsour1}) and (\ref{eq:cylsour2}) connect the 
radiation field inside the hemisphere with the source 
function with correction factors $\CI_\pm$ being given by:
\beq
\CI_+(\tau,\tau',\mu)&=&\left\{ \begin{array}{ll} 
0, & \mbox{if} \quad t\geq r+r', \\
1 , & \mbox{if} \quad t\leq r'-r, \\
\CI , & \mbox{if} \quad r'-r\leq t \leq r+r',  
\end{array} \right.   \\
\CI_-(\tau,\tau',\mu)&=&\left\{ \begin{array}{cl} 
0, & \mbox{if} \quad t\geq r+r', \\
( r'/r )^2, & \mbox{if} \quad t\leq r-r', \\
\CI , & \mbox{if} \quad r-r' \leq t \leq r+r',  
\end{array} \right.   
\eeq 
where 
\bez
t=\sqrt{1-\mu^2} |\tau-\tau'| / \mu , \quad
r=\sqrt{\taut^2-\tau^2} ,   \quad
r'=\sqrt{\taut^2-\tau'^2} , 
\eez
and
\beq
\CI&=&\frac{1}{\pi r^2}\left[ r^2 \phi_*+ 
r'^2 \cos^{-1}\left( \frac{r'^2-r^2+t^2}{2r't}\right) 
 -rt\sin \phi_* \right] , \nonumber \\
\phi_*&=&\cos^{-1}\left( \frac{r^2-r'^2+t^2}{2rt}\right) .  
\eeq
 
The emerging flux through the  base of the hemisphere  is 
 $\mu\pi I^-(\tau=0,x,\mu)$.  
The flux through the   curved hemisphere surface   is given by 
equations~(\ref{eq:cylflux}) and (\ref{eq:cylfl2}), 
where $\CE_\pm$ are (note that $\taut=\taur$ and $\mu>0$):
\beq
\CE_+(\tau,\tau',\mu)&=&\left\{ \begin{array}{ll} 
0,                   & \mbox{if} \quad t\geq r+r', \\
2\pi\mu\tau/\taut , & \mbox{if} \quad t\leq r'-r, \\
C_+ ,                & \mbox{if} \quad r'-r \leq t \leq r+r',  
\end{array} \right.   \\
\CE_-(\tau,\tau',\mu)&=&\left\{ \begin{array}{ll} 
0,    & \mbox{if} \quad t\geq r+r' ,\; \mbox{or}\quad t \leq r-r', \\
C_- , & \mbox{if} \quad r-r' \leq t \leq r+r',  
\end{array} \right.  
\eeq
\beq
C_\pm&=&2\left[ \pm\mu\phi_\pm\tau/\taut + \sqrt{1-\mu^2}\sqrt{1-(\tau/\taut)^2} 
\sin \phi_\pm \right] , \nonumber \\
\phi_+&=&  \left\{ \begin{array}{ll} 
\phi_*, & \mbox{if} \quad \mu\geq \sqrt{1-(\tau/\taut)^2} , \\
\min\{ \pi/2, \phi_*\} , & \mbox{if} \quad \mu < \sqrt{1-(\tau/\taut)^2} ,
\end{array} \right.   \\
\phi_-&=&\phi_* . \nonumber
\eeq

\subsection{Isotropic Source Function Approximation}
\label{sect:isosf}

In some applications where  high accuracy is not needed and we are not 
interested in the polarization of the radiation, we can substantially
reduce the computing time assuming that the source function, 
${S}_k(\tau,x,\mu)$, is isotropic and homogeneous for 
scattering orders $k\geq  2$. The accuracy of this approximation is discussed in 
\S~\ref{sect:appr}. The approximation works because for the optically thin 
coronae photons scattered 
more than a few times are almost isotropic and 
are distributed almost homogeneously throughout the medium. 
In this case, the iteration procedure starting from the second 
scattering order can be written as follows: 
\beq 
{S}_{k+1}(x)&=&x^2\int_0^{\infty}
 \frac{\rmd x_1}{x_1^2} {R}(x,x_1){J}_{k}(x_1) , \\
{J}_{k+1}(x)&=&{S}_{k+1}(x) P_{\rm J}(x), \quad k\geq 1, 
\eeq
where 
\be
 {J}_{k}(x)=\frac{1}{\taut}\int_0^{\taut}\rmd \tau \frac{1}{2} 
\int_{-1}^{1}\rmd \mu  {I}_k(\tau,x,\mu) 
\ee
is the   intensity  averaged over optical depth and angles, and 
\be
{R}(x,x_1)= \int_{0}^{1}\rmd \mu \int_{0}^{1}\rmd \mu_1 
\left[ {R}_{11}(x,\mu;x_1,\mu_1) + {R}_{11}(x,\mu;x_1,-\mu_1) \right]
\ee 
is the angle averaged  Compton redistribution function. 
The quantity, $P_{\rm J}(x)$,  can be obtained from equations~(\ref{eq:cylsour1}) 
and (\ref{eq:cylsour2}): 
\beq \label{eq:pes}
P_{\rm J}(x)&=&\frac{1}{\taut}\int_0^{\taut}\rmd \tau \frac{1}{2} 
\int_{0}^{1}\frac{\rmd \mu}{\mu} \left[ 
 \int_{0}^{\tau} \rmd \tau'  \exp\left\{-  
\int_{\tau'}^{\tau}\sigma(\tau'',x,\mu)
\frac{\rmd \tau''}{\mu}\right\}   \CI_+(\tau,\tau',\mu) \right. \nonumber \\ 
&+&\left.
  \int_{\tau}^{\taut}   \rmd \tau' 
\exp\left\{- \int_{\tau}^{\tau'}\sigma(\tau'',x,-\mu)
\frac{\rmd \tau''}{\mu}\right\}   \CI_-(\tau,\tau',\mu) \right] .
\eeq 
In  the case of slab geometry, where $\CI_\pm=1$, and at photon energies, 
$x< 1$,  where 
pair  production is not important and hence $\sigma(\tau,x,\mu)=\sigmacs(x)$,
this integral can be computed analytically: 
\be
P_{\rm J}(x)=\taut \frac{1}{\taux}\left[ 1-\frac{1}{\taux}\left( \frac{1}{2} 
-E_3(\taux)\right) \right]  , 
\ee
where $\taux=\taut\sigmacs(x)$ is the frequency dependent optical depth  
and $E_3$ is the exponential integral  of the third order. 
For $\taux\ll 1$, we have
\be
P_{\rm J}(x) \sim 
 \taut \frac{1}{2} \left( -\ln\taux + \frac{3}{2} -\gamma_E \right) ,
\ee
where  $\gamma_E=0.577216...$ is the Euler's constant. 

The emergent fluxes through the top and bottom of the cylinder, and through the
base of the hemisphere, and from the slab, 
 can be found from equations~(\ref{eq:cylsour1}) and (\ref{eq:cylsour2}) 
 where the source functions, ${S}_k$, can now be taken out from the integrals.
Corresponding $\mu$-dependent multiplicative factors  similar to $P_{\rm J}(x)$ 
 can be computed before   the iteration procedure.  Similarly, the emergent 
fluxes through the side of the cylinder, and through the curved surface of the 
hemisphere can be found using equations~(\ref{eq:cylflux}) and 
(\ref{eq:cylfl2}).  
 
\section{The  Balance Equations} \label{sec:baleq}

\subsection{The Energy Balance}  \label{sec:energy}

It is common for problems where  pair production is important that the 
luminosities appear in the form of  dimensionless compactnesses.     
We define the compactnesses in the following way: 

1. For an active region (a coronal cylinder or a  coronal hemisphere) 
 
A  dissipation compactness, $\ldiss \equiv (\Ldiss H/R^2)(\sigmat/\me c^3)$, 
characterizes the dissipation 
with $\Ldiss$ being the power providing uniform heating of the active region, 
and $R$ being the radius of cylinder, and $H$ its height. For hemispheres we 
have $H=R$.
The soft compactness $\ls\equiv  (\Ls H/R^2)(\sigmat/\me c^3)$ characterizes 
the soft (reprocessed plus internally dissipated) luminosity from the cold disk 
entering the active region; 
$\lc\equiv  (\Lc H/R^2)(\sigmat/\me c^3)$
is the coronal compactness corresponding to
the total luminosity of Compton  scattered radiation and radiation emitted
 in the corona. 

2. For a plane-parallel slab corona

A local dissipation compactness is defined as 
 $\ldiss \equiv (\Ldiss /H )( \sigmat / \me c^3 )$ with $\Ldiss$ 
being the power providing uniform heating of a  cubic volume of size $H$ 
in the slab.  Similar definitions hold for $\ls$ and $\lc$. 

The soft compactness, $\ls$, consist of two parts, $\ldisk$ and $g\lrepr$, 
where $\ldisk$ is the compactness of the power that is internally dissipated 
in the cold disk and that enters the corona,  
and $\lrepr$ is the compactness of the power reprocessed by the cold disk. 
The parameter $g$ is the fraction of the radiation reprocessed and
reflected from the cold disk which enters the active region. 
Introducing the parameter $d\equiv\ldisk/\ldiss$, 
we can write the energy balance equation for the cold phase as 
\be \label{eq:ls}
\ls=g\lrepr+d\ldiss . 
\ee
If all power dissipates in the corona then $d=0$. 
The total coronal compactness, $\lc$, is the sum of the compactness
dissipated in the corona, $\ldiss$, and the part of the soft and reflected 
compactnesses which is scattered in corona: 
\be \label{eq:lc}
\lc=\ldiss+p_{\rm sc}(\ls+g\lrefl) . 
\ee
This equation represents the energy balance of the hot phase (i.e. the corona). 
Here  $p_{\rm sc}$ is the probability of scattering in the corona  for disk 
photons entering the base of the active region  ($p_{\rm sc}$ is a geometry 
dependent function of $\taut$); 
and $\lrefl$ is the compactness  reflected from the cold disk.  
Introducing the integrated disk albedo, $a$, which is the fraction of the 
incident luminosity reflected by the cold disk,  and the anisotropy parameter, 
$\eta$, which is the fraction of all coronal radiation (Comptonized, annihilation, 
bremsstrahlung, and double Compton radiation) that is incident  on the cold 
disk, we can write: 
\be \label{eq:lrefl}
\lrefl=a\eta\lc, \quad \lrepr=(1-a)\eta\lc .
\ee
The  equations (\ref{eq:ls})-(\ref{eq:lrefl}) can easily be solved  
for the  ratios $\lc/\ldiss$ and $\ls/\ldiss$: 
\beq \label{eq:lcldiss}
\frac{\lc}{\ldiss}&=&  \frac{1+p_{\rm sc}d}{1-p_{\rm sc}g\eta},  \\ 
\frac{\ls}{\ldiss}&=& \frac{g\eta (1-a)+d(1-p_{\rm sc}g\eta a)}
  {1-p_{\rm sc}g\eta} .
\label{eq:lsldiss}
\eeq
 Defining  the amplification factor, $A \equiv \lc/\ls$, we obtain
\be \label{eq:ampli}
A=\frac{1+p_{\rm sc}d}{g\eta(1-a)+d(1-p_{\rm sc}g\eta a)} .
\ee  
If all power dissipates in the corona  we have  $A=1/g\eta(1-a)$. 

We now show how to compute the parameters entering the energy balance equations
from  the solution of radiative transfer. Let us define the partial flux 
emergent from the corona after $k$ scatterings as (note that $\mu>0$):
\beq
\tilde{F}_k^+(x,\mu)&=& \mu  \tilde{I}_k^+(\tau=\taut,x,\mu), \nonumber\\
\tilde{F}_k^-(x,\mu)&=& \mu  \tilde{I}_k^-(\tau=0,x,\mu) , 
\eeq 
for slab geometry. For cylinder geometry, the expressions are the 
following:  
\beq
\tilde{F}_k^+(x,\mu)&\equiv&\tilde{F}_k^{\rm side,+}(x,\mu) +
\mu\pi \tilde{I}_k^+(\tau=\taut,x,\mu), \nonumber\\
\tilde{F}_k^-(x,\mu)&\equiv&\tilde{F}_k^{\rm side,-}(x,\mu) +
\mu\pi \tilde{I}_k^-(\tau=0,x,\mu) . 
\eeq
Analogous expressions hold for hemisphere geometry, but 
the upward flux is just
$\tilde{F}_k^+(x,\mu)=\tilde{F}_k^{\rm side,+}(x,\mu)$. 
Let us also define $\tilde{F}_0^{{\rm in},+}$ as the emergent flux of 
 unscattered soft and reflected radiation entering the corona (note that 
$\tilde{F}_0^{{\rm in},-}=0$). The emergent flux of unscattered radiation 
emitted in the corona is $\tilde{F}_0^{\pm}$. 
The total emergent coronal flux is then given by  $\displaystyle 
\tilde{F}^\pm_{\rm c}=
\tilde{F}_0^{\pm}+\sum_{k=1}^\infty \tilde{F}_k^\pm $, 
and the total  emergent  flux is $\tilde{F}^\pm=
\tilde{F}_0^{{\rm in},\pm}+\tilde{F}^\pm_{\rm c}$.
The total reflected radiation can be found by convolving of 
Green's matrix for Compton reflection with the total flux incident on
the cold disk:
\be \label{eq:refl}
\tilde{I}_{\refl}(x,\mu)=\int_x^\infty \rmd x_1\int_0^1 \frac{\rmd\mu_1}{\mu_1}
\h{G}(x,\mu;x_1,\mu_1)\tilde{F}^-(x_1,\mu_1) . 
\ee
The integrated disk albedo, $a$, is the ratio of the reflected flux to 
the total flux incident on the cold disk:   
\be \label{eq:albedo}
a= \displaystyle \frac{\int_0^\infty \rmd x 
\int_0^1 {I}_{\refl}(x,\mu) \mu\rmd\mu }
{ \int_0^\infty \rmd x \int_0^1  {F}^-(x,\mu)  \rmd \mu } . 
\ee
The anisotropy parameter, $\eta$, is given by 
\be \label{eq:aniso}
\eta= \displaystyle \frac{\int_0^\infty \rmd x \int_0^1 {F}^-_{\rm c}
(x,\mu) \rmd\mu }{ \int_0^\infty \rmd x \int_0^1  \left[ 
{F}^+_{\rm c}(x,\mu) + {F}^-_{\rm c}(x,\mu) \right] \rmd \mu } , 
\ee
and the scattering probability, $p_{\rm sc}$, for slab geometry is given by 
\be \label{eq:psc}
p_{\rm sc}=1- \displaystyle\frac{\int_0^\infty \rmd x \int_0^1 
{F}^{{\rm in},+}_0(x,\mu) \rmd\mu }
{ \int_0^\infty \rmd x \int_0^1  {I}_{\rm in}(x,\mu) \mu \rmd \mu } , 
\ee
with  similar expressions for hemisphere and cylinder geometry, 
but a factor $\pi$ should then be introduced in the denominator.  
In the calculations, $d$ is specified and $g$ is determined in advance, 
while $a$, $\eta$, and $p_{\rm sc}$ are calculated from the radiative 
transfer results  using expressions (\ref{eq:albedo})--(\ref{eq:psc}), and
the amplification factor, $A$, is given by equation (\ref{eq:ampli}).

The sum of  $\ldiss$ and $\ldisk$ can be written as the sum of the total
upward emergent flux and the total downward emergent flux that does not
reenter the corona: 
\be \label{eq:ldissflux}
\ldiss+\ldisk= 2\pi \int_0^\infty \rmd x \int_0^1 \left[
F^+(x,\mu) +(1-g) F^-(x,\mu) \right] \rmd\mu .
\ee
The actual value of  $\ldiss$ is determined by pair balance, but does not 
influence the energy balance.

\subsection{The Pair Balance}    \label{sec:pair}

For the range of temperatures of interest, $\Theta < 2$, particle-particle 
and particle-photon pair production is negligible
compared to photon-photon pair production.
The pair annihilation rate, $\dot{n}_{\rm ann}\cm^{-3}$,
is uniform throughout the corona, while the pair production
rate, $\dot{n}_{\g\g}(\tau)\cm^{-3}$, 
depends on the radiation 
field inside the medium being largest at the center where the photon 
density is largest and smallest at the boundaries.  
In pair balance, the pair annihilation rate, $\dot{n}_{\rm ann}$,
 is equal to the volume-averaged pair production rate: 
\be
\dot{n}_{\rm ann}=\ovl{\dot{n}}_{\g\g}, 
\ee
where
\be 
\ovl{\dot{n}}_{\g\g}= \left\{ \begin{array}{ll} 
\displaystyle\frac{1}{\taut}\int_0^{\taut} \dot{n}_{\g\g}(\tau) \rmd \tau , & 
\mbox{for slabs and cylinders} ,  \\
&  \\
\displaystyle\frac{3}{2\taut}\int_0^{\taut} \left[ 1- (\tau/\taut)^2 \right] 
\dot{n}_{\g\g}(\tau) \rmd \tau , &
\mbox{for hemispheres} . 
\end{array} \right. 
\ee
Here the extra factor in
the integrand for hemispheres accounts for the decreasing volume of horizontal 
layers with height.
The expressions for  $\dot{n}_{\rm ann}$ and $\dot{n}_{\g\g}(\tau)$
are given in  Appendices~\ref{sec:ggrate} and \ref{sec:annrate}.

\subsection{The Iteration Procedure}    \label{sec:iter}

For a given coronal geometry (slab,   cylinder, or hemisphere atop of the cold 
disk) 
only two  parameters, $\ldiss$ (or, alternatively, 
$\Theta$) and $\Tbb$, uniquely specify the simulations if all power is 
dissipated in corona ($d=0$). 
For $d\not= 0$, two more parameters, $d$ and $\Tdisk$, should be specified.  
The equilibrium state satisfies the energy and pair balance equations 
coupled with the radiative transfer. To find the solution we make use of an 
iterative  procedure. 
For a given $\Theta$, we compute the Compton redistribution 
matrix and the cross section, the coronal emissivities (annihilation and 
bremsstrahlung), and guess the initial values for the $\ldiss$ and $\taut$. 
The reflected spectrum and the pair production absorption coefficient are set 
to be zero. Initial values  for the parameters in the energy balance equations 
are $a=0$, $\eta=1/2$, $p_{\rm sc}=0$. 
We  also compute the amplification factor $A$, and the 
soft compactnesses, $\lrepr$ and $\ldisk$. 
We then normalize the  incident black body radiation using equations 
(\ref{eq:bbnorm}) or (\ref{eq:bbnormcy}) and thus obtain the incident 
spectrum, $\tilde{I}_{\rm in}(x,\mu)$, from equations (\ref{eq:iinsl}) 
or (\ref{eq:cylbou}). 
 Solving the radiative transfer  by expansion in scattering 
orders we find the radiation field inside the medium as well as the emergent 
fluxes. We then compute the rate of pair production and the absorption 
coefficient, the double Compton emissivity, 
the reflected spectrum,  and the albedo, $a$,  the  
parameters $\eta$, and $p_{\rm sc}$, and the amplification factor,  $A$.

By comparing   $\ovl{\dot{n}}_{\g\g}$ with $\dot{n}_{\rm ann}$ 
we calculate a new imposed dissipation 
compactness $\ldiss^{\rm new}=\ldiss^{\rm old} \sqrt{\dot{n}_{\rm ann}/
\ovl{\dot{n}}_{\g\g}}$. 
Comparing the calculated amplification factor $A^{\rm new}$ with $A^{\rm old}$ we choose 
the new optical depth, $\taut$, to be smaller than the old  $\taut$ if 
$A^{\rm new}>A^{\rm old}$,  and larger if  $A^{\rm new}<A^{\rm old}$. 
The change, $\Delta\taut$,   decreases by a factor of two for  
on each iteration when the sign $\Delta\taut$   changes. 
 After that we start the next iteration by again solving   the 
 radiative transfer. 

The number of iterations needed to achieve an accuracy better than $1\%$ in all 
equations is about 10.  
On a Sparc 20 a typical simulation takes about 5 minutes for 6 angular points,
7 spatial zones, and 80 frequency points.  The isotropic source function 
approximation (see \S~\ref{sect:isosf}) reduces the 
computing time for   solving  the radiative transfer problem by an order 
of magnitude.

\section{Comparison with Other Codes} \label{sec:compar}    

\subsection{Comparison with Non-Linear Monte-Carlo Results} 

\begin{figure}
\plotone{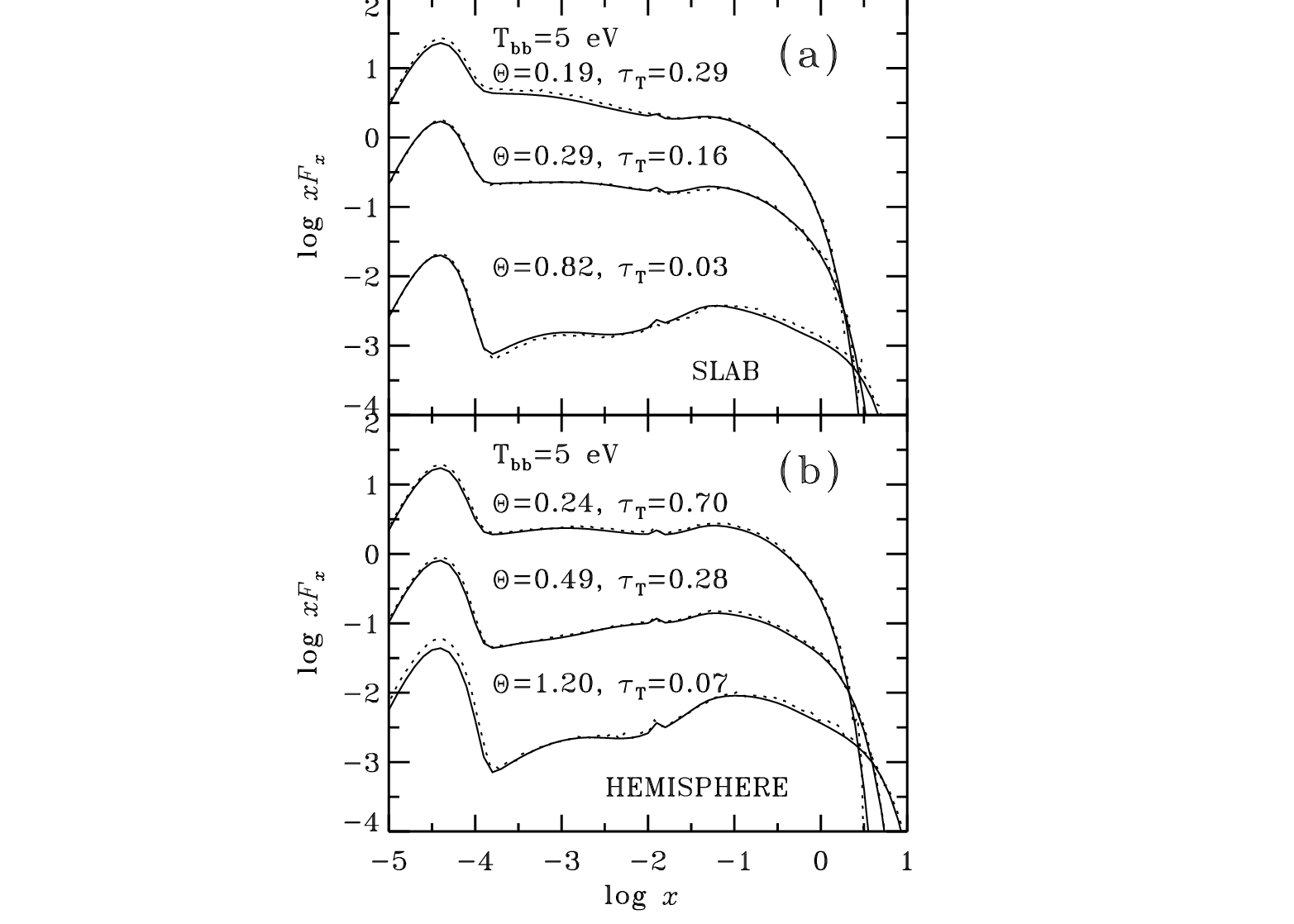}
\caption{ (a) Emergent flux, $xF_x$ (arbitrary units), from  slab pair coronae 
as function of dimensionless photon energy, $x\equiv h\nu/\me c^2$, 
for the parameters in Table~1. 
The spectra are averaged over  viewing 
angles $0.6<\mu<0.9$. {\em Solid} and {\em dotted curves}: results using the  
ISM code and NLMC code, respectively.  
(b)  Same as (a), but for  hemisphere coronae.  
}
\label{fig:1}
\end{figure}

We compare our calculations based on the iterative scattering method (ISM)
with the corresponding results using the 
Non-Linear Monte-Carlo (NLMC) code  
by Stern (see Stern et al. 1995a). We made 3 test runs each for
slabs and for hemispheres   atop of the cold disk.  We assume that all power 
dissipates in the corona ($d=0$). The parameter $g$ is calculated 
 in the iteration procedure assuming that only the radiation reprocessed below 
the base of the active region actually reenters the active region.  
 The results are given in Table~1 and are shown in Figure~\ref{fig:1}. 
We find that for a given $\Theta$, the optical depth, $\taut$, 
is almost the same for both codes with the largest difference being 
about 5-8 per cent at small $\taut$.
The difference in the derived compactnesses, $\ldiss$, is less than 20 per cent 
for slab geometry. In the case of  hemispheres, $\ldiss$ 
differs by 20 per cent at large $\taut$,
and is a factor two smaller at small $\taut$ due to our approximate 
treatment of the radiative transfer in hemispheres. The procedure of averaging
 the radiation field
over the horizontal layers artificially increases the photon density in the 
active region causing  pair balance to be reached at smaller compactnesses. 
For large $\taut$, the difference
is smaller due to smaller boundary effects. The differences can  also be
due to our assumed homogeneity of the corona, rather than using a number of
zones as in the NLMC method. 
As shown in Figure~\ref{fig:1}, both codes give quite similar spectral shapes 
for the emerging radiation for both types of geometries. 
The differences in $\ldiss$ can be considered as small (at least for the slab 
case) if we remember that 
$\Theta$ and $\taut$   depend  rather weakly on $\ldiss$. Thus, if 
we fix $\ldiss$ instead of $\Theta$, the differences in $\Theta$ and 
$\taut$ will be about 2 per cent. 

\subsection{The Accuracy of some Useful Approximations} 
\label{sect:appr}

A number of approximations can be used to decrease the  time needed to 
compute  the redistribution function for Compton scattering. One is
the isotropic scattering approximation in the electron rest frame 
(see eq.~[\ref{eq:8piQ}] in Appendix~\ref{sec:iso}). 
We found that for the  mildly relativistic temperatures considered here 
the Comptonization spectra computed in this approximation is very accurate 
at small energies ($x<\Theta$) but have deficits of photons at higher energies. 
Solving the pair balance in this approximation gives $\ldiss$ a factor 
of 6 larger for large $\taut$ (small $\Theta$), and a factor of 2 larger at 
small $\taut$ (large $\Theta$) (see Table~1 where this approximation is
denoted ISOSCAT1). 

\begin{figure}
\plotone{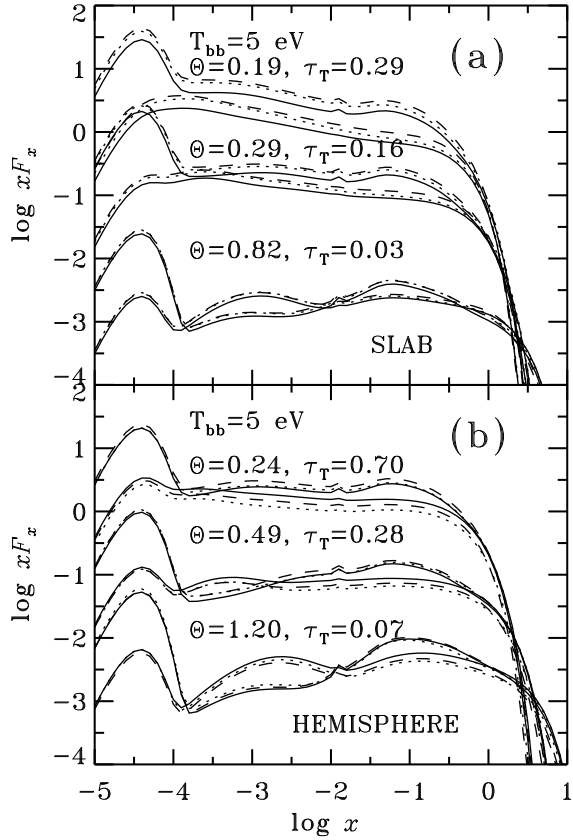}
\caption{Same as Figure~\protect\ref{fig:1}    but for two viewing angles, 
$\mu=0.11$ and $\mu=0.89$.  The face-on spectra ($\mu=0.89$) can be 
identified by their more prominent black body component.
 All curves represent results from the ISM code.   
Results using the exact scattering kernel are shown by 
 {\em solid curves}.  {\em Dashed curves} represent results 
 assuming  isotropic scattering in the 
electron rest frame  (ISOSCAT2, see Appendix A2).  {\em Dotted curves}  
correspond to the isotropic source function approximation (ISOSF).  
Notice that the emergent spectra for the different approximations 
are normalized to the   corresponding  value of $\ldiss$ in Table~1. 
The spectral fluxes in a given direction differ slightly, but the spectral 
shapes are almost identical. 
} 
\label{fig:2}
\end{figure}

To improve the high energy behavior of the Comptonized spectra, we used
the redistribution function from equation~(\ref{eq:8piQ}) and the exact 
value for $\g_*$ from  equation~(\ref{eq:gam*}). This approximation works 
much better, but still produces 
deficit of high energy photons. We found the resulting $\ldiss$ to be 
10-30 per cent too large 
(see Table~1 where this approximation is denoted ISOSCAT2). Spectra
in this approximation (dashed curves) are compared with exact results (solid 
curves) in Figure~\ref{fig:2}. 
At low energies 
($x<\Theta$) the spectra are almost identical, but the approximate spectra fall 
more rapidly at larger
energies forcing the compactness to increase in order to satisfy the pair
balance. Due to the small contribution of photons with $x\sim 1$ to the total 
energy balance, the optical depth differs by less than 1 per cent from the 
exact calculations.
We conclude that this approximation is useful for modeling spectra
of mildly relativistic pair plasmas.

In many works, the pair production rate is computed assuming an 
isotropic radiation field. The isotropic pair production cross section, 
$R_{\g\g}^{\rm iso}(xx_1)$ (see Appendix~\ref{sec:symmpr}), 
is much easier to compute than the angle-dependent pair production
cross section, $R_{\g\g}(xx_1,\mu,\mu_1)$  (see e.g. Coppi 
\& Blandford 1990). 
In the problem at hand, the radiation field is 
strongly anisotropic. We investigated the errors caused by making the
radiation field isotropic before computing the pair production rate
 (see Table~1 where this approximation is denoted ISORAD). 
The resulting $\ldiss$ are 
systematically lower than those obtained using the exact angle-dependent 
$R_{\g\g}$, 
because of a higher pair production rate for the isotropic case. 
The effect is  smaller at large $\taut$ where the radiation field is more 
isotropic.  
Changes in the pair production rate   have no influence on the
energy balance and consequently does not change the computed spectra and 
optical depths. It does, however, influence the value of the compactness.

In \S~\ref{sect:isosf} we presented a method of solving the radiative 
transfer equation when the source function is assumed to be isotropic starting 
from the second scattering order. 
The overall spectral shapes in this approximation (dotted curves) are quite   
similar to the exact spectra (solid curves) in Figure~\ref{fig:2}.
The ISOSF approximation underestimates the flux in the ``edge-on'' direction
for   hemisphere geometry due to the artificial isotropization of the radiation 
field. The differences in optical depth are 
negligible for slab geometry, but become about 3-10 per cent for hemispheres.  
The resulting $\ldiss$ are 10-50 per cent too large  (see   Table~1 where this
approximation is denoted by ISOSF).

\subsection{Comparison with HM93}

\begin{figure}
\plotone{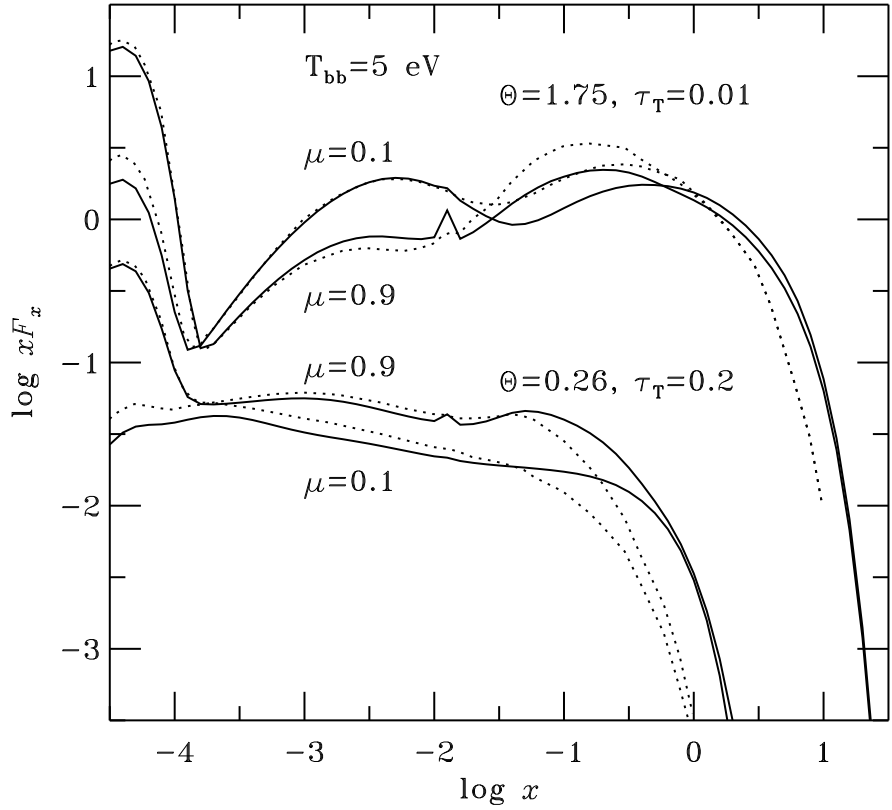}
\caption{Emergent flux, $xF_x$ (arbitrary units), from  slab pair coronae as 
function of dimensionless photon energy, $x\equiv h\nu/\me c^2$. 
{\em Solid curves}: results using the ISM code for $\mu=0.11$ and $\mu=0.89$. 
{\em Dotted curves}: results from Figures~4a,c in HM93 for $\mu=0.15$ and 
$\mu=0.95$. 
}
\label{fig:3}
\end{figure}

In Figure~\ref{fig:3} we compare our calculations with the results from 
Fig.4a-c in HM93 for   slab geometry. 
In the  low energy band they are almost identical, but at high energies 
($x>\Theta$) the spectra of HM93 have too sharp cutoffs reflecting their 
use of an {\it ad hoc}
exponential cutoff $\exp (-x/\Theta)$. The actual cutoff  energy 
is approximately $x\approx 2\Theta$, and  the cutoff is not a true 
exponential, but rather reflects the thermal Compton scattering kernel
and the distribution of the emergent photons over the scattering orders.
The spectral indices in the $2-18$ keV range, $\a218$, 
are very close (see Table~2) for small $\Theta$, but differs at larger $\Theta$.
The reason  probably lies in the treatment of the reflection from
the cold disk. HM93 computed the reflection using the  Monte-Carlo method 
assuming
isotropic incident flux. For large $\Theta$, the {\it anisotropy break} 
(see Stern et al. 1995b) occurs
in the $2-18$ keV range. The flux incident on the cold disk is thus  very 
anisotropic having its maximum along  the normal to the disk. Making the flux
isotropic by angle averaging artificially increases the flux along the plane 
of the disk, which has a larger probability for reflection (Matt 1993; Poutanen
et al. 1996). 
The contribution of the reflection component to the total flux increases 
making the 2-18 keV spectra flatter. This explains the difference in  $\a218$
at large $\Theta$.

The $\Theta$-$\taut$ relation  obtained by our code agrees with calculations of 
HM93 to within a percent or two for small $\taut\approx 0.01$.
The difference increases with increasing $\taut$ and at our largest 
$\taut$ = 0.37 the $\Theta$ of HM93 is about 10 per cent 
too large due to our spectral differences and the  corresponding influence on 
the  energy balance.
The anisotropy factor, $\eta$, (see Fig. 2a in HM93) agrees very well up to $\taut$ = 0.1. 
Above that the $\eta$ of HM93 is slightly too small, becoming 0.02 smaller 
 at $\taut$ =0.37 most likely due to our $\Theta$ being 10 per cent smaller.
 Our albedo, $a$, is smaller (0.13 instead of 0.16)
at small $\taut$ (see Fig. 2b in HM93). Above   $\taut=0.05$, our $a$ is 0.01 
larger. The differences are likely due to our use of a fully relativistic 
and anisotropic treatment of the reflection. 

The comparison of compactnesses is not so easy. First of all, HM93 give the 
values for the Compton compactness $\lc$ which is related to the total 
dissipation compactness as $\ldiss=(1-p_{\rm sc}\eta)\lc$, where $p_{\rm sc}$ is the 
probability of scattering in corona (see \S~\ref{sec:energy}). 
Second, they define the compactness to be factor of $\pi$ larger than our 
definition.  Thus, in Table~2 we give the  corrected  
 $\ldiss= (1-p_{\rm sc}\eta)\lc/\pi$, instead of the 
original values of $\lc$  from Table~1 in HM93.  
For small $\taut$, our compactnesses are a factor  5 smaller, but for 
$\taut=0.2$ our compactness is larger than corresponding compactnesses of HM93.
The approximate estimates by HM93 of the 
pair producing photon density inside the slab (see Appendix B in HM93) and 
 the prescription for the Comptonized spectra (Zdziarski 1985) used in 
the pair balance calculations are responsible for the remaining differences. 

\subsection{Comparison of Comptonized Spectra with Analytical Formulae}

\begin{figure}
\plotone{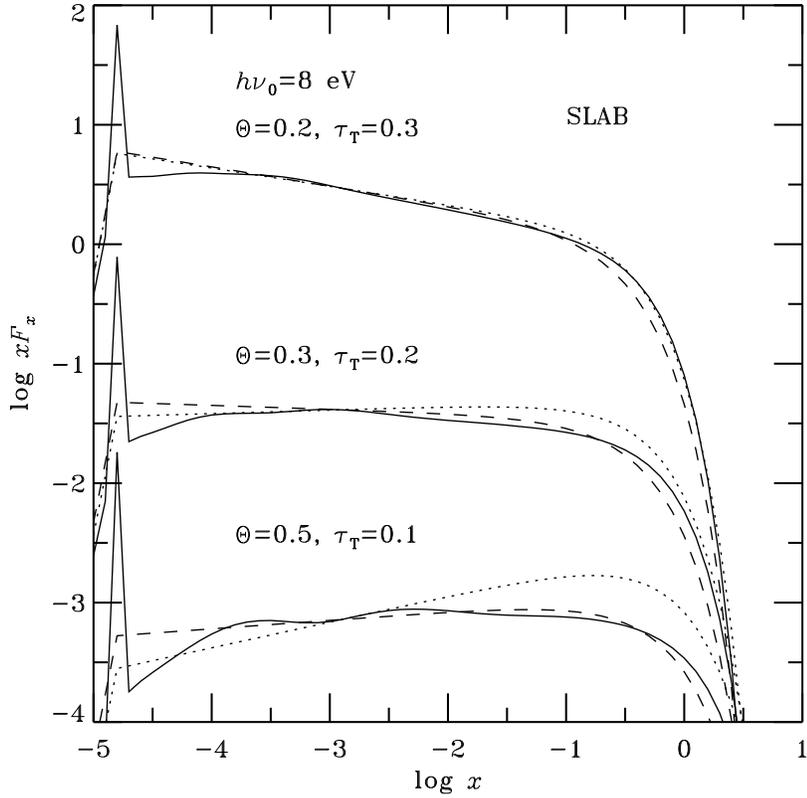}
\caption{Angle averaged Comptonized spectra, $xF_x$ (arbitrary units),
from  slab coronae  as 
function of dimensionless photon energy, $x\equiv h\nu/\me c^2$.  
Soft monochromatic photons with $h\nu_0=8$ eV are assumed to be
incident on the lower boundary of the slab.
{\em Solid curves}: results using the ISM code;  
{\em dotted curves}: spectra using the analytical formulae~(35) and (44) 
from Titarchuk (1994); {\em dashed curves}: 
spectra using the analytical formulae~(9) and (10) from Hua \& Titarchuk (1995).
No pair or energy balance is imposed.  
}
\label{fig:4}
\end{figure}

We compare the angle averaged Comptonized spectra computed using 
the ISM code with the analytical formulae for thermal Comptonization from 
the papers by Titarchuk (1994,  his eqs. [35] and [44]) 
and Hua \& Titarchuk (1995, their eqs. [9] and [10]). 
We consider  monochromatic incident photons with $h\nu_0=8$ eV on the 
lower boundary of the slab. The only process which is taken into account
is Comptonization.  No pair or energy balance is imposed.  
Calculations for three different optical depths and 
temperatures are presented in Figure~\ref{fig:4}. All three cases
correspond to  regime 2 in Hua \& Titarchuk (1995). 
We find that the Hua \& Titarchuk 
formulae  rather well  represent the general spectral 
behavior, but give systematically fewer photons in the high energy tail. 
The Titarchuk formulae, however, give a very good description of the 
spectra for relatively small temperatures ($\Theta<0.2$), while they produce 
too many photons in the Wien bump for larger temperatures. 
In all these cases we should, however,  remember that spectra below the 
anisotropy break as well as the high energy tail 
have quite different behavior at different viewing angles. 
The analytical formulae 
do not provide this angular dependence of the spectrum, 
and  are therefore quite limited in practice.

\subsection{Polarization Properties}

In this section, we compare our calculations of the degree of polarization 
of the radiation emerging from the Compton scattering slab-corona with some  
earlier results by others. 
We assume here that the cold disk emits semi-isotropic  unpolarized radiation. 
We define the degree of polarization as $p=(Q/I) \;100 \%$. The polarization
 is positive when the electric vector is predominantly parallel  to the normal 
to the slab. The behavior of the total polarization is affected by both 
the Compton scattering radiation from the hot corona, and by reflected 
radiation from the cold disk.  
 
\begin{figure}
\plotone{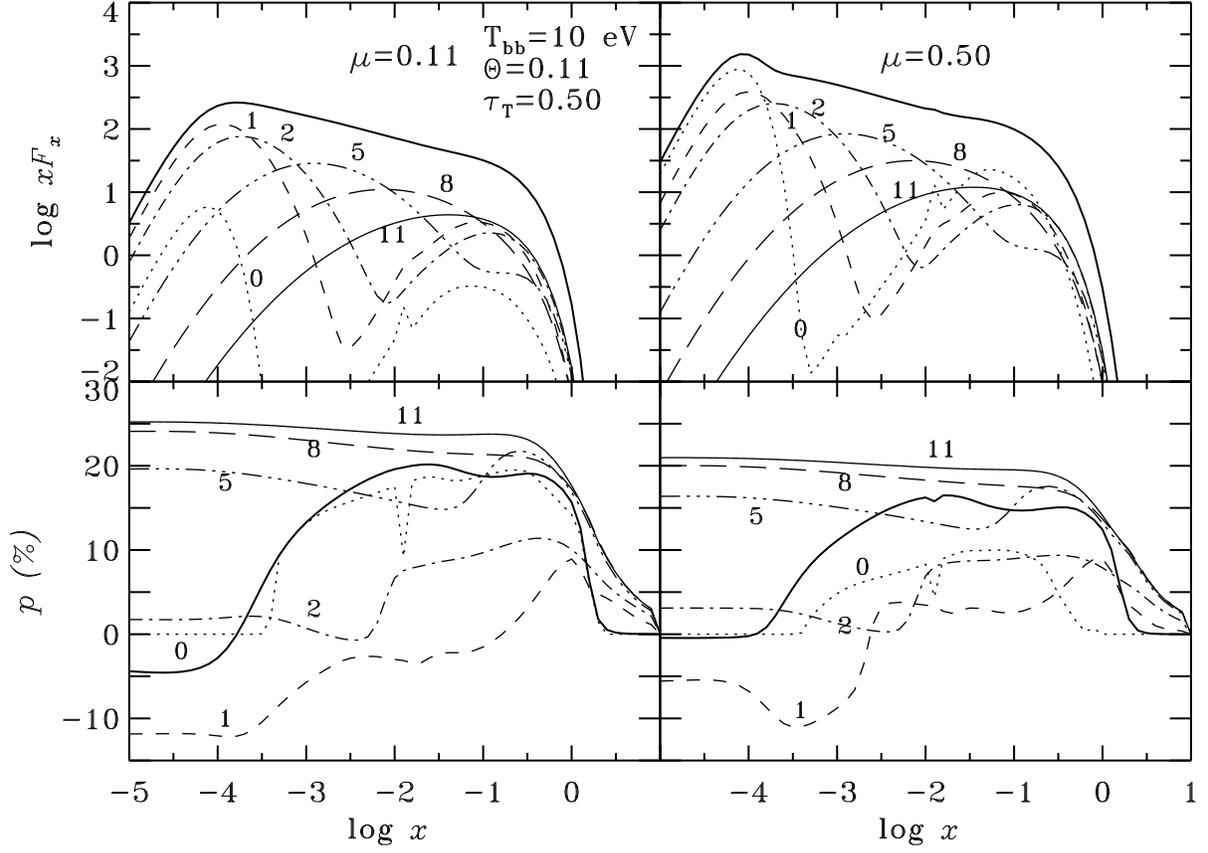}
\caption{Emergent flux, $xF_x$, from  slab coronae (upper panels) and 
the degree of polarization, $p$ (lower panels) as 
function of dimensionless photon energy, $x\equiv h\nu/\me c^2$, 
for $\Theta=0.11$, $\;\taut=0.50$, $\;\Tbb=10$ eV for two viewing angles 
$\mu=0.11$ and $\mu=0.50$. 
No pair or energy balance is imposed.  
{\em Thick solid curves}: the overall spectra and polarization. 
The contribution from some of the   scattering orders is also shown. 
The labels show the order of scattering. 
The zeroth ({\em dotted}) component consists of the unpolarized blackbody 
disk radiation as well as  the radiation reflected from the cold disk.  
The scattered components consist of multiple scattered blackbody radiation and
multiple scattered reflected radiation with the latter being centered around 
$x\sim 0.1$. 
}
\label{fig:5}
\end{figure}

Sunyaev \& Titarchuk (1985) 
calculated the  polarization of the Comptonized radiation  from  a slab. 
The angular and polarization structure of the radiation field were 
obtained with an iteration procedure based on an expansion in scattering orders 
using the Rayleigh matrix. 
  The frequency dependence of the intensity was obtained solving the Kompaneets 
equation (see Sunyaev \& Titarchuk 1980), which is a diffusion equation 
in frequency space. The distribution of photons over escape time and thus the 
 relation between the   scattering order and 
frequency were obtained by solving the diffusion equation in physical space.  
This approach has a few shortcomings. First, the diffusion approximation in 
optical depth (i.e. physical space) is not valid for the optically thin coronae 
considered here. 
Second, for the case of a hot electron gas, the frequency redistribution 
cannot be considered as diffusion due to the large frequency shift in each 
scattering. 
And last, for high electron temperatures ($\Theta\simgreat 0.1$) 
and/or large photon energy  ($x\simgreat 0.1$),  
the polarization properties of the exact Compton redistribution 
matrix  are quite different from those of the Rayleigh matrix (Poutanen 
\& Vilhu 1993).   
Thus, even for  quite small electron temperature, $\Theta=0.11$, and 
relatively large optical depth, $\taut=0.5$, the maximum polarization of the
hard radiation computed
by this method is $p\approx 50\%$ (see Fig.~8 in Sunyaev \& Titarchuk 1985, 
note also that their $\tau_0=\taut/2$), compared to $p\approx 25\%$ in our 
 calculations (see our Fig.~\ref{fig:5}).   
The differences become much larger for smaller $\taut$ and/or larger $\Theta$.

\begin{figure}
\plotone{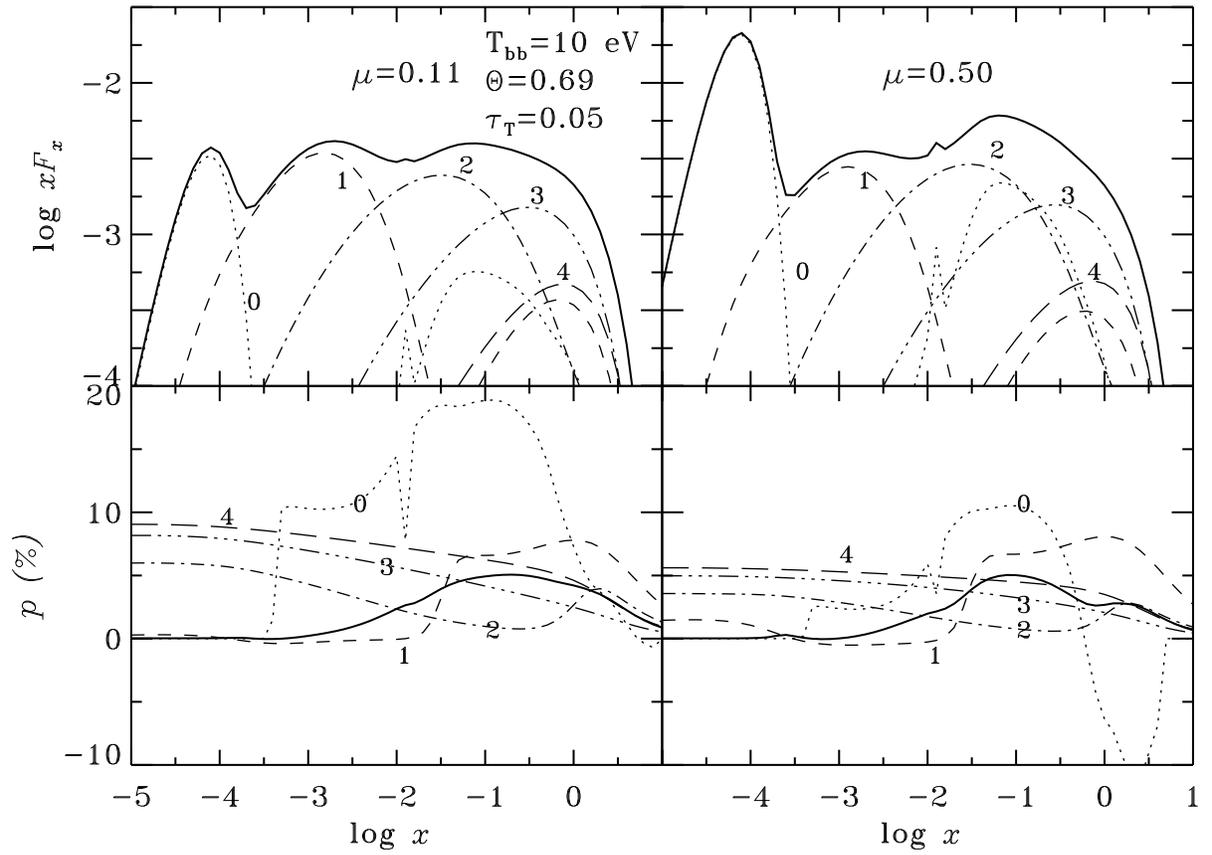}
\caption{Same as Figure~\protect\ref{fig:5} but for $\taut=0.05$ and 
$\Theta=0.69$.
}
\label{fig:6}
\end{figure}

Following the same idea of separating the polarization structure from 
the frequency redistribution,  Haardt \& Matt (1993) computed the degree of 
polarization from an optically thin hot slab-corona
applying the method of Haardt (1993) to obtain the spectra.  
Using an iterative scattering scheme similar to ours they avoided the diffusion 
approximation both in frequency and optical depth space. They, however,  still 
used the Rayleigh matrix. To compare the results,   
we have chosen the same parameters  as in  Figure~2 in Haardt \& Matt (1993). 
The blackbody temperature is taken  to be, $\Tbb=10$ eV, 
the  optical depth is $\taut=0.5$ and $0.05$. 
We show the results of our computations in Figures~\ref{fig:5} and \ref{fig:6}.   
We present the results for two viewing angles, $\mu=0.11$ and $\mu=0.50$. 
The degree of polarization 
is  zero in the direction  normal to the slab due to  symmetry. 
It is clearly seen that the degree of polarization is much too large
in Haardt \& Matt (1993). They found the maximum polarization to be 
$p\approx45\%$ for $\taut=0.5$ and $p\approx 33\%$ for $\taut=0.05$, 
while we obtained $p\approx 25\%$ and $p\approx 5\%$, respectively 
(thick solid curves in Figs.~\ref{fig:5} and \ref{fig:6}).

Below, we discuss the polarization  properties both of the Comptonized radiation  
from the hot corona and of the radiation  reflected from the cold slab. 
In order to better  see   the contribution to the overall polarization 
from different scattering orders 
we also show the spectra for individual scattering orders as well as the total 
spectrum in the upper panels of Figures~\ref{fig:5} and \ref{fig:6}.  
 
First we consider the polarization caused by scattering in the hot corona.
The degree of polarization  increases with the number
of scatterings reaching  its asymptotic value after few scatterings. The 
asymptotic value depends on  optical depth,  temperature and  zenith angle. 
At  energies close to $\me c^2$, the Klein-Nishina corrections start to be 
important decreasing the  polarization. 
As the electron temperature increases 
the polarization decreases (Poutanen \& Vilhu 1993).  
At a given frequency the contribution from
the higher scattering orders (with larger polarization) becomes smaller 
also    decreasing  the polarization. Due to these two reasons  the 
polarization in Figure~\ref{fig:5}  ($\taut=0.5$) is larger than 
in  Figure~\ref{fig:6} ($\taut=0.05$). 
The polarization of a given scattering order first decreases toward higher 
energies but then increases due to the contribution from the scattered
reflected component. 
The decrease is caused by the fact that the largest energy shift is obtained  in
backward scatterings which do not produce additional polarization.

The reflection component   contributes to the overall spectra at
  energies   $x\approx 0.01-1$. 
When the flux incident on the cold disk   is nearly isotropic, which is the case 
for  optically thin corona, the polarization of the reflected component is 
positive  at 
$x \simless 1$. It is maximal in edge-on directions, and is zero in the normal
direction (see Poutanen et al. 1996 for a discussion of the polarization 
properties of   Compton-reflected  radiation from the cold slab). 
The  polarization in  the direction close to the normal decreases at higher 
energies, changing   sign at $x\sim 0.6$. 
The flux reflected close to the normal direction cuts off  at energies 
above $x\sim 1$. At directions closer to the plane of the slab, the 
 cutoff is slower, and is determined  by the cutoff of the incident spectrum. 
The polarization  has a sharp feature at 6.4 keV due to the contribution from 
the unpolarized  fluorescent iron line.  

For $\taut=0.5$ (Fig.~\ref{fig:5}) the polarization of the reflected radiation 
(and that of the reflected component scattered once or twice in the corona) is 
smaller 
than the polarization of the component produced by Compton scatterings in 
hot corona. This causes the polarization to decrease at $x> 0.1$. On 
the other hand for $\taut=0.05$ (Fig.~\ref{fig:6}) the reflected component has 
significantly larger polarization than the scattered component resulting in a
smoothly increasing  polarization in the energy interval  
from  $x\approx 0.01$ up to $x\sim 0.1$. The decrease of the polarization
 of the reflected component at higher energies and the change of sign 
at $x\sim 0.6$  for $\mu=0.5$ cause the drop in the total  polarization  
at $x\sim 1$.

 Future observations of
X-ray polarization by {\em Spectrum-X}-$\gamma$ satellite (Kaaret et al. 1992)  
can be a powerful tool for determining  the physical conditions 
and, probably, the geometry of the X-ray emitting region in AGN and X-ray 
binaries.   
We note here that the degree of polarization   for hemisphere and 
cylinder geometries is smaller than for the slab case. 
We can conclude that if   small  polarization in the X-rays will be observed
this could argue  for   Comptonization models where the temperature of the 
electron gas is large and/or for   models where the geometry of the corona 
is not  slab-like.

\section{Summary and Conclusions}  \label{sec:concl}  

We have described a versatile code based on the iterative
scattering method (ISM) to accurately solve the 
radiative transfer and Comptonization 
in a two-phase disk-corona models for active galactic nuclei and X-ray
binaries.

The code has several attractive features some of them being unique to
this code:

1) The radiative transfer is fully angle-dependent, and
one can easily determine the outgoing spectrum in any direction to high
accuracy at any photon energy. The outgoing spectrum in
a given direction may differ greatly from the angle-averaged spectrum.
 
2) The radiative transfer is valid for both nonrelativistic and
relativistic temperatures.

3) The radiative transfer is polarized.

4) Most important radiation processes in hot thermal plasmas
including Compton scattering,
photon-photon pair production, pair annihilation, bremsstrahlung, and
double Compton scattering are taken into account. The latter
two were not important for the parameters of the test cases
we considered in this paper.
 
5) The corona can be in energy and/or pair balance.

6) The corona can either be a pure pair corona, or one can include a
background plasma.

7) The reflection by the cold disk is exactly treated using a 
reflection matrix that, in particular, accounts for the 
full angular dependence of the incident radiation.

8) The ISM code has been extensively tested against a Non-Linear Monte Carlo 
(NLMC) code (Stern et al. 1995b) finding very good agreement.  

9) The ISM code is an order of magnitude faster than the NLMC code. 
The ISM code also allows for an easier 
determination of the spectral fall-off at photon energies above $k\Te$.
The ISM code also gives more accurate
emerging spectra at a given viewing angle as compared to the NLMC code 
where one must average over a range of viewing angles in order to improve the
photon statistics. 
  
10) Various approximations for the radiative transfer/Comptonization can be
used in order to improve computing efficiency.
Quite accurate results can 
be obtained if one assumes the Compton scattering source
function to be isotropic. The gain in computing efficiency
is then an order of magnitude. A typical run on a Sparc 20 
then takes about 40 s for  three angular, 80
frequency, and 7 spatial gridpoints. 
 
There are, however, some limitations:

1) The iterative scattering method converges only for small optical depths. 
 For small temperatures ($\Theta<0.1$) and large optical depths
($\tau>1$), the round off errors become
large due to the large number of scatterings  and the accuracy of the 
results decreases. The maximum allowed  $\taut$ depends on the temperature 
and is approximately $1$ for 
$\Theta=0.1$, and 1.5 for $\Theta=0.5$ in the slab case 
 and   $2-3$ in the case of hemisphere geometry. 
 This is not much of a limitation at 
temperatures above about 100 keV, as the optical depths needed
to explain observed X-ray spectral indices in  AGN are necessarily less 
than unity.

2) The ISM code is one-dimensional. The ISM code can, however,
be applied to quasi-1D radiative transfer
in   two-dimensional active regions with cylinder or hemisphere geometry 
atop or elevated above the cold disk.
 The NLMC code can, in principle, treat arbitrary geometries.

3) The corona is uniform in temperature and density.
The Nonlinear Monte Carlo code has been used (Stern et al. 1995b)
to show that the differences in temperature and density across the corona
is at most a factor 2. In principle, one could divide the corona 
into a few zones, and    solve the radiative transfer 
equation using the appropriate Compton redistribution functions and 
cross sections corresponding to the  temperature in each zone. The pair and 
energy balance equations should then be solved in each  zone separately. 
 This, however, is  much more time consuming than
considering a homogeneous corona. The resulting spectra from homogeneous 
and inhomogeneous corona are very similar. 

4) The ISM code treats steady radiative transfer.  The NLMC code,
on the other hand, can treat time dependent situations.
  
There are several possible applications of the ISM code:

1) Results of earlier work using various approximations can be checked,
and the validity of the approximations tested. In particular,
we find that the full Compton redistribution matrix, rather
than the Rayleigh scattering matrix,
must be used in order to obtain accurate polarized 
X-ray spectra.
 
2) The validity of published analytical fits for
angle-averaged Comptonized spectra can be checked.
The results of the ISM code can be used to obtain analytical fits 
of spectra as function of viewing angle and for different geometries.

3) The ISM code has great potential for modeling 
X-ray and $\gamma$-ray spectra from active galactic nuclei 
and X-ray binaries. The ISM code has already
been used together with a NLMC code to interpret the 
statistics of observed X-ray spectral
indices and compactnesses from Seyfert 1 galaxies (Stern et al. 1995b).
The anisotropy of outgoing spectra is an important ingredient
in this interpretation, which could not have been done using
angle-averaged model spectra.

4) The ISM code is highly suitable for inclusion
in spectral fitting software such as XSPEC, whereupon
observed X-ray spectra can be modeled. Such modeling of a number of
sources will  appear in forthcoming work.

5) The ability of the ISM code
to treat polarized radiative transfer makes it a powerful
tool for interpreting future observations of
X-ray polarization from, e.g., the {\it Spectrum-X-$\gamma$} and 
{\it INTEGRAL} satellites.

\acknowledgments 
The authors thank Boris Stern for stimulating discussions. 
This research was  supported by
grants and a postdoctoral fellowship (J.P.) 
from the Swedish Natural Science Research Council. 

\appendix
\section{Rates of the Physical Processes} \label{sec:compsc}

\subsection{The Electron Scattering Source Function and the  
Compton Redistribution  Matrix} \label{sec:sourcefunc}

The thermal electron scattering source function, $\tilde{S}(\tau,x,\mu)$, 
for an azimuth-independent 
radiation field accounting only for linear polarization    
can be expressed in terms of the 
azimuth-averaged Compton redistribution matrix, $\h{R}(x,\mu;x_1,\mu_1)$,
as:
\be \label{eq:sour}
\tilde{S}(\tau,x,\mu)=
 x^2\int_0^{\infty} \frac{\rmd x_1}{x_1^2}\int_{-1}^{1} \rmd\mu_1  
 \h{R}(x,\mu;x_1,\mu_1) \tilde{I}(\tau,x_1,\mu_1) \, ,
\ee
Here, $\h{R}$  is the azimuth-averaged  product of 
two rotational matrices, $\h{L}$,  and the thermal 
Compton redistribution matrix, $\h{C}(x,x_1,\costh)$ (Poutanen \& Vilhu 1993; the 
hat identifies $\h{R}$, $\h{L}$, and $\h{C}$ as matrices, the tilde identifies
$\tilde{S}$ and $\tilde{I}$ as  vectors):
\be 
\h{R}(x,\mu;x_1,\mu_1)=\int_0^{2\pi}\rmd \varphi 
\h{L}(-\chi) \h{C}(x,x_1,\costh) \h{L}(\chi_1) .
\ee
 In general, it is  $4\times 4$ matrix.
The rotational matrices are given by the following expression 
(see, e.g., Chandrasekhar 1960): 
\be
\h{L}(\chi) =\left( \begin{array}{c} 1\\0\\0\\0 \end{array}
\begin{array}{c}  0 \\
 \begin{array}{r}\cos(2\chi)\\  -\sin(2\chi) \end{array} \\0 \end{array}
\begin{array}{c}  0 \\
 \begin{array}{r}\sin(2\chi)\\  \cos(2\chi) \end{array} \\0 \end{array}
 \begin{array}{c} 0\\0\\0\\1 \end{array}  
 \right) . 
\ee 
 Due to   azimuthal symmetry and the absence of  
circular polarization we consider only the $2\times 2$ matrix in the  
upper left corner of the general matrix:
\be \label{eq:csmre}
\h{R}= \left( \begin{array}{cc}  
 {R}_{11}  &  {R}_{12}\\
 {R}_{21}  &  {R}_{22} 
\end{array} \right) .
\ee  
The elements of this  matrix  are: 
\beq 
 R_{11}&=&\int C \rmd\varphi, \nonumber  \\
 R_{12}&=&\int  C_{\rm I}\cos 2\chi_1 \rmd\varphi,  \nonumber\\
 R_{21}&=&\int  C_{\rm I}\cos 2\chi \rmd\varphi,  \nonumber\\
 R_{22}&=&\int  [  C_+\cos 2(\chi-\chi_1) + C_-\cos 2(\chi+\chi_1)]
 \rmd\varphi, 
\eeq
where $C_\pm =(C_{\rm Q}\pm C_{\rm U})/2$,  and
the cosines are given by  
\be
 \cos\chi=\frac{\mu_1-\mu\costh}{\sinth\sqrt{1-\mu^2}}, \quad 
\cos\chi_1=\frac{\mu-\mu_1\costh}{\sinth\sqrt{1-\mu^2_1}}, 
\ee
with $\costh=\mu\mu_1+ \sqrt{1-\mu^2}\sqrt{1-\mu^2_1}\cos\varphi$. 
Finally, the functions $C, C_{\rm I}, C_{\rm Q}$, and $C_{\rm U}$ are four 
of the five functions forming the thermal Compton redistribution matrix, 
$\h{C}(x,x_1,\costh)$.  This matrix is given by a single integral 
over the electron energy distribution $f(\g)$ and the Compton
redistribution matrix, $\h{C}^m(x,x_1,\costh,\g)$, 
for an isotropic {\it monoenergetic} electron gas with Lorentz factor $\g$
(Nagirner \& Poutanen 1993):
\beq\label{eq:rmaxw}
\lefteqn{ \h{C}(x,x_1,\costh)=\left( \begin{array}{llll}
C  & C_{\rm I} &  0  & 0 \\
C_{\rm I} & C_{\rm Q} &  0  & 0 \\
 0  &  0  & C_{\rm U} & 0 \\
 0  &  0  &  0  & C_{\rm V} 
\end{array} \right)  }
\nonumber\\
&=&
\frac{3}{8}
\int_{\g_*}^\infty f(\g) 
\h{C}^m(x,x_1,\costh,\g)
\rmd\g 
=
\frac{3}{8}
\int_{\g_*}^\infty f(\g)\rmd\g 
\left( \begin{array}{llll}
 C^m  & C_{\rm I}^m &  0  & 0 \\
C_{\rm I}^m & C_{\rm Q}^m &  0  & 0 \\
 0  &  0  & C_{\rm U}^m & 0 \\
 0  &  0  &  0  & C_{\rm V}^m 
\end{array} \right) ,   
\eeq
where
\beq \label{eq:gam*}
 \gamma_*\equiv [x-x_1+Q(1+2/q)^{1/2}]/2,  \qquad q\equiv 
xx_1(1-\costh) , \qquad Q^2 \equiv (x-x_1)^2+2q .\nonumber
\eeq
The function $C^m$ is the scalar redistribution function derived by 
Jones (1968) (see also Coppi \& Blandford 1990). 
We use the following expressions to calculate the five functions that
enter $\h{C}^m(x,x_1,\costh,\g)$ in  equation~(\ref{eq:rmaxw}):  
\beq
  C_{}^m&=&C_{\rm a}^m+C_{\rm b}^m,\nonumber \\ 
  C_{\rm I}^m&=&C_{\rm a}^m+C_{\rm c}^m,  \nonumber\\
 C_{\rm U}^m&=&\frac{2}{Q}++2\frac{u-Q}{rq}\left[ \frac
{u-Q}{rq}(2Q+u)-4\right] +\frac{2u}{vq} +2R_{\rm c}^m, \\
C_{\rm Q}^m&=&C_{\rm U}^m+C_{\rm a}^m, \nonumber \\
 C_{\rm V}^m&=&C_{\rm b}^m-qC_{\rm a}^m, \nonumber
\eeq
where
\beq  \label{eq:Rabc}
 C_{\rm a}^m&=&u\frac{(u^2-Q^2)(u^2+5v)}{2q^2v^3}+u\frac{Q^2}{q^2v^2}, \nonumber\\ 
 C_{\rm b}^m&=&\frac{2}{Q}+\frac{u}{v}\left(1-\frac{2}{q}\right) , \\ 
 C_{\rm c}^m&=&\frac{u}{vq}\left( \frac{u^2-Q^2}{rq}-2\right) ,\nonumber
\eeq
and 
\beq
 u=a_1-a=(x+x_1)(2\g +x_1-x)/(a+a_1),  \qquad v=aa_1,    \\
 a^2=(\g -x)^2+r, \qquad a_1^2=(\g +x_1)^2+r, \qquad r=(1+\costh)/(1-\costh) .  
\nonumber 
\eeq
The  normalized relativistic Maxwellian distribution is given by 
\be
f(\g)=\frac{e^{-\g/\Theta}}{4\pi\Theta K_2(1/\Theta)} , 
\ee
which gives the  density of particles in the dimensionless momentum volume, 
$4\pi z^2dz$, normalized to unity. 
Here, $K_n$ is the modified Bessel function of  second kind of order $n$, and
$z\equiv \sqrt{\gamma^2-1}$.
Methods for computing   the integrals when averaging $\h{C}^m(x,x_1,\costh,\g)$ 
over a relativistic Maxwellian electron
distribution are given in Poutanen~(1994).  If the electron temperature is 
not very high ($\Theta < 1$) we can use  Gauss-Laguerre quadrature.

\subsection{The Compton Redistribution Function for  
Isotropic Scattering in the Electron Rest Frame}
\label{sec:iso}

A very simple expression for the redistribution function, $C^m$, can be derived 
if we assume that the characteristic photon energy in the electron rest frame 
is small  $x_1\gamma \ll 1$, i.e. the Thomson limit. 
The scattering in this limit can be assumed to be coherent in the electron
rest frame.
We also assume that the scattering is isotropic in the rest frame. 
Such simplifications give a correctly  
normalized redistribution function at energies $x\ll\Theta$  (see 
eq.~[\ref{eq:norm}]), whose shape slightly differs from the exact one. 
The function $C^m$ in that approximation becomes (Arutyunyan \& Nikogosyan
 1980)
\be \label{eq:43q}
C^m=\frac{4}{3Q} \; ,   
\ee 
being non-zero when  
\be \label{eq:coq}
2q (\g^2-1) \geq (x_1-x)^2.
\ee  
Integration over $\costh$ between the limits defined by
 equation~(\ref{eq:coq}) gives the 
 angle-averaged  redistribution function  (Rybicki \& Lightman 1979):  
\be \label{eq:iiso}
C^m(x,x_1,\g)=\int  C^m \sinth\rmd \theta=
\frac{ 4}{3xx_1}
\left[ x+x_1 - \frac{\g}{z}\left| x-x_1\right| \right] ,
\ee 
where $x/x_1 \in [(\g-z)^2, (\g+z)^2]$.  
For Maxwellian electrons the redistribution function is given by
\be \label{eq:8piQ}
C(x,x_1,\costh)=\frac{1}{8\pi Q}\frac{e^{-\g_*/\Theta}}{K_2(1/\Theta)}, 
\ee 
where  $\g_*=Q/\sqrt{2q}$. Equation~(\ref{eq:8piQ}) gives very good 
approximation to the exact redistribution function
 for mildly relativistic temperatures and  
$x\ll\Theta$.  Notice also that $xC(x,x_1,\costh)$ is a function of 
 the ratio $x/x_1$.  In Table~1 this approximation is called ISOSCAT1.
Even better agreement  with the exact redistribution function is obtained if 
$\g_*$ from  equation~(\ref{eq:gam*}) is used in  equation~(\ref{eq:iiso}). 
 We call this approximation ISOSCAT2 in Table~1. 
The accuracy of these approximations is discussed in \S~\ref{sect:appr}.

\subsection{The Thermally Averaged Compton Scattering  Cross Section} 

The Compton scattering cross section averaged over a relativistic Maxwellian electron 
distribution can be written as an  integral over the electron energy:
\beq 
\sigmacs(x)&=&\frac{3\sigmat}{16x^2\Theta K_2(1/\Theta)}
\int_{1}^{\infty} e^{-\g/\Theta}\left[ \left( x\gamma+\frac{9}{2}+
\frac{2}{x}\g\right) \ln\frac{1+2x(\g+z)}{1+2x(\g-z)} - 2xz \right.  \nonumber\\
& +&\left. z \left( x-\frac{2}{x}\right) \ln(1+4x\g+4x^2) + 
\frac{4x^2z(\g+x)}{1+4x\g+4x^2}- 
2\int_{x(\g-z)}^{x(\g+z)}\ln(1+2\xi)\frac{\rmd\xi}{\xi} \right] \rmd \gamma .
\eeq 
Making the substitutions $\g=1+\Theta\exp(-2t)$ on the  interval $[1,1+\Theta]$, 
and  $\g=1+\Theta(1+t)$ on the  interval $[1+\Theta, \infty)$, 
and applying 10-points Gauss-Laguerre quadrature formula we achieve 
an accuracy better than 0.02 per cent.
 
In limiting cases, the thermal cross section can be computed using   
simple expressions (Gould 1982; Svensson 1982; Nagirner \& Poutanen 1994): 
\beq 
\sigmacs(x)&=&\frac{3\sigmat}{8x^2} \left[ 4+ \left( x -2 -\frac{2}{x}\right)
 \ln(1+2x) +
  \frac{2x^2(1+x)}{(1+2x)^2} \right] , \qquad \Theta \ll 1,  \\
\sigmacs(x)&=&\frac{3\sigmat}{16x\Theta} \left( \frac{1}{2}-\g_E +
\ln 4x\Theta \right) , 
\qquad \qquad \Theta\gg 1,\; x\Theta\gg 1, \\
\sigmacs(x)&=&\frac{\sigmat}{K_2(1/\Theta)} \sum_{n=0}^{\infty} (-2x)^n a_n 
 K_{n+2}(1/\Theta) , \qquad \qquad x\Theta\ll 1, 
\eeq 
where  
\be 
a_n=\frac{3}{8} \left[ n+2+\frac{2}{n+1}+\frac{8}{n+2} -\frac{16}{n+3} 
\right] . 
\ee 
These simple approximations can be  used to check the correctness of the 
thermal cross section routine and to estimate the numerical accuracy. 

\subsection{Symmetry Properties and a Normalization Condition} \label{sec:symmcs}

The azimuth and thermally averaged functions ${R}_{ij}$ have symmetry
properties which can be exploited to simplify the radiative transfer
equation, to reduce the time needed to calculate all elements of the
redistribution matrix, and to check the accuracy of the calculations:  

1.   Frequency symmetry  
\be 
 {R}_{ij}(x,\mu;x_1,\mu_1)e^{-x_1/\Theta} = {R}_{ij}(x_1,\mu;x,\mu_1)
e^{-x/\Theta}, \quad  i,j=1,2 \, ,   
\ee
which follows from microscopic detailed balance between states $x$ and $x_1$
when the photons and the electrons have a Wien and a Maxwellian distribution,
respectively (Pomraning 1973; M\'esz\'aros \& Bussard 1986). 
The exponential factors represent 
the Wien distribution, while the photon phase space factors have
 been absorbed in the definition of ${R}_{ij}$.

2.  Angular  symmetry  
\beq 
 {R}_{ij}(x,\mu;x_1,\mu_1)&=& {R}_{ji}(x,\mu_1;x_1,\mu) ,   \nonumber \\
 {R}_{ij}(x,\mu;x_1,\mu_1)&=& {R}_{ij}(x,-\mu;x_1,-\mu_1) , \quad
i,j=1,2\;.
\eeq
These angular symmetries follow directly from the fact that the scattering
process depends on the scattering angle between the in and outgoing photons, and
not on their angle cosines, $\mu$ and $\mu_1$, separately.

The thermal Compton scattering cross section, $\sigmacs(x)$, 
and the scalar redistribution function, ${R}_{11}(x_1,\mu_1;x,\mu)$,
i.e. element $11$ of the thermal redistribution matrix, are related through
a normalization condition (see e.g. Pomraning 1973; Nagirner \& Poutanen 1994):
\be \label{eq:norm}
\frac{\sigmacs(x)}{\sigmat} =\frac{1}{x}\int_0^\infty x_1 \rmd x_1 
\int_{0}^{1} \rmd \mu\int_{0}^{1} \rmd \mu_1
\left[  {R}_{11}(x_1,\mu_1;x,\mu) +
 {R}_{11}(x_1,\mu_1;x,-\mu) \right] \, . 
\ee
Analogous integrations of  ${R}_{12}$ and  ${R}_{21}$ gives zero on the left
hand side  due to the requirement that the polarization is zero ($Q=0$) for
an isotropic radiation field.
These relations can be used to check the accuracy of the calculation of the
redistribution matrix, and to estimate the quality of the frequency and the 
angular discretization.

\subsection{The Photon-Photon Pair Production Rate} \label{sec:ggrate}  

For azimuth-independent (i.e. axisymmetric) photon distributions, 
the rate of photon-photon 
pair production, $\dot{n}_{\g\g} \cm^{-3}\sec^{-1}$,  neglecting polarization 
is given by integrals over dimensionless photon energy, $\rmd x$, and solid
angle, $2 \pi \rmd\mu$, as follows: 
\be 
\dot{n}_{\g\g}(\tau)=\frac{1}{2} \frac{2\pi}{\me c^2} \int_0^\infty  \frac{\rmd x}{x} 
\int_{-1}^1  I(\tau,x,\mu)    \alpha_{\g\g}(\tau,x,\mu) \rmd \mu . 
\ee 
Here, $I(\tau,x,\mu)\rmd x/(c\me c^2x)$ is the number density of photons of 
energy $x$ in the interval $\rmd x$ per steradian traveling in the direction 
$\mu$, and $\alpha_{\g\g} \cm^{-1}$ is  the absorption coefficient
due to photon-photon pair production. The factor 1/2 is due to both interacting
species being photons. The absorption coefficient is given by  another
integral over the target photon energy, $\rmd x_1$, and solid angle, 
$\rmd\varphi \rmd\mu_1$, 
\be
\alpha_{\g\g}(\tau,x,\mu)=\frac{\re^2}{\me c^3}  \int_0^\infty 
\frac{\rmd x_1}{x_1} \int_{-1}^1 
R_{\g\g}(x,\mu;x_1,\mu_1)   I(\tau,x_1,\mu_1)  \rmd \mu_1  ,  
\ee
where $\re$ is the classical electron radius, 
$R_{\g\g}(x,\mu;x_1,\mu_1)$ is the dimensionless, azimuth-integrated
pair production cross section, 
\be \label{eq:pairfunc}
R_{\g\g}(x,\mu;x_1,\mu_1)=2 \int_{\varphi_{\min}}^{\pi} 
s_{\g\g}(\omega)(1-\cos \theta) \rmd\varphi ,
\ee
the $(1- \cos \theta)$ factor is  discussed in, e.g., Weaver (1976),
and $s_{\g\g}(\omega)$ is the dimensionless photon-photon pair production  
cross section (Jauch \& Rohrlich 1976)
\be \label{eq:paircross}
s_{\g\g}(\omega) \equiv
\frac{\sigma_{\g\g}(\omega)}{\re^2} = \frac{\pi}{\omega}\left[ 
\left( 2+ \frac{2}{\omega}-\frac{1}{\omega^2}\right) {\rm cosh}^{-1} 
\sqrt{\omega} - \left( 1+\frac{1}{\omega}\right) 
\sqrt{1-\frac{1}{\omega} } \right] .
\ee 
Here, $\omega \equiv x_{cm}^2 = xx_1(1-\cos \theta)/2$, where $x_{cm}$ is
the photon energy in the center-of-momentum frame,
and $\theta$ is the interaction angle related to
other cosines as 
$\cos \theta =\mu\mu_1+\sqrt{1-\mu^2}\sqrt{1-\mu_1^2} \cos\varphi$.
Using these relations, the threshold condition for pair production, 
$x_{cm} > 1$, can be written
as a constraint on $\cos \theta$ or on $\cos \varphi$ 
giving a minimum allowed value for $\varphi$: 
$\cos\varphi_{\min}=(1-\mu\mu_1-2/xx_1)/
\left( \sqrt{1-\mu^2}\sqrt{1-\mu_1^2}\right) $.
The factor 2 in equation~(\ref{eq:pairfunc}) comes from the integration
range originally being $\varphi_{\min}$
to $2 \pi - \varphi_{\min}$, and the integrand being an even function of $\varphi$
around $\varphi = \pi$. 
The axisymmetric pair production rate was previously considered by 
Stepney \& Guilbert (1983) who chose $x_{cm}$ as integration variable
instead of $\varphi$. Their rate is a factor 2 too large as pointed out 
by Kusunose (1987).

\subsection{The Thermal Pair Annihilation Rate} \label{sec:annrate}

For a relativistic Maxwellian electron (and positron) distribution, the
pair annihilation reaction rate, $\dot{n}_{\rm ann} \cm^{-3}\sec^{-1}$, can 
be written as a one-parameter ($\Theta$) single integral (Weaver 1976).
Svensson (1982) made a simple fit to that integral accurate to within 2 per cent: 
\be 
\dot{n}_{\rm ann}=n_-n_+c\re^2\frac{\pi}{1+2\Theta^2/\ln(1.3+2\eta_E\Theta)}, 
\ee
where $\eta_E\approx 0.5615$.

\subsection{Pair Annihilation Emissivity}

The  emissivity  due to thermal pair annihilation, $\ep_{\rm ann} (x, \Theta) 
\rmd x \ergs \cm^{-3} \sec^{-1} \sr^{-1}$, in an energy interval $\rmd x$ 
can be written using detailed balance
arguments in terms of the 
pair production cross section in the following form (Svensson 1983): 
\be \label{eq:annemi}
\epsilon_{\rm ann}(x,\Theta)  =n_-n_+\re^2 \me c^3
\frac{xe^{-x/\Theta}}{2\pi\Theta K_2^2(1/\Theta)} 
\int_1^{\infty}  \omega  s_{\g\g}(\omega) e^{-\omega /x\Theta} \rmd\omega , 
\ee 
where $s_{\g\g}(\omega)$ is given by  equation~(\ref{eq:paircross}).
Simple analytical fits 
for the one-parameter ($x\Theta$) integral in equation~(\ref{eq:annemi}) 
accurate to within 0.04 per cent are given by Svensson, Larsson, \& Poutanen
(1996). 

\subsection{Symmetry Properties of the Azimuth-integrated Pair Production
Cross Section} 
\label{sec:symmpr}

The azimuth-integrated cross section, $R_{\g\g}$, obey energy and angular 
symmetry relations, which are useful in reducing the computing time:
\beq
R_{\g\g}(x,\mu;x_1,\mu_1)&=&R_{\g\g}(xx_1,\mu,\mu_1), \nonumber \\
R_{\g\g}(x,\mu;x_1,\mu_1)&=&R_{\g\g}(x,\mu_1;x_1,\mu) , \\
R_{\g\g}(x,\mu;x_1,\mu_1)&=&R_{\g\g}(x,-\mu;x_1,-\mu_1) . \nonumber 
\eeq 
These symmetry relations follow directly from equation~(\ref{eq:pairfunc}),
the definition of $\omega$, and the relations for $\cos \theta$ and 
$\cos \varphi_{\min}$.  
To check the accuracy of our calculations  we  integrate the 
azimuth-integrated cross section   
over one cosine angle and average over the second in order to obtain
the fully solid angle-integrated cross section for the isotropic case
(which is a well known function first computed by Gould and Schr\'eder 1967): 
\be \label{eq:riso}
R_{\g\g}^{\rm iso}(xx_1)=
\frac{1}{2}\int_{-1}^1 \rmd \mu \int_{-1}^1 \rmd \mu_1 
R_{\g\g}(x,\mu;x_1,\mu_1) =
\int_0^1 \rmd \mu \int_0^1 \rmd \mu_1 
\left[ R_{\g\g}(x,\mu;x_1,\mu_1) +  R_{\g\g}(x,\mu;x_1,-\mu_1) \right].
\ee
Here we have used the third symmetry property above.
The angle-averaged function, $\bar \phi (xx_1)$ in Gould \& Schr\'eder (1967)
is related to our $R_{\g\g}^{\rm iso}(xx_1)$ through
$\bar \phi (xx_1)= (xx_1)^2 R_{\g\g}^{\rm iso}(xx_1) / 8 \pi^2$. 
The angle-averaged cross section, $R(xx_1)$  in Coppi \& Blandford (1990),
is related to our  $R_{\g\g}^{\rm iso}(xx_1)$ through
$R(xx_1)= c\re^2 R_{\g\g}^{\rm iso}(xx_1)/ (4 \pi)$.
Coppi \& Blandford (1990) give a useful fit for $R(xx_1)$ accurate to within
7 per cent for all $xx_1$. We find our $R_{\g\g}^{\rm iso}(xx_1)$ computed using
equation~(\ref{eq:riso}) to
typically be accurate to within 2 per cent.
 
One can show that the annihilation emissivity can be written as
an integral, not over the pair production cross section, but over
the angle-integrated pair production cross section:
\be \label{eq:annnor}
\epsilon_{\rm ann}(x,\Theta)=n_-n_+\re^2 \me c^3
\frac{x^3e^{-x/\Theta}}{16\pi^2\Theta^2 K_2^2(1/\Theta)} 
\int_0^{\infty} x_1^2 R_{\g\g}^{\rm iso}(xx_1)  e^{-x_1/\Theta} \rmd x_1 . 
\ee 
By numerically computing this integral using our computed 
$R_{\g\g}^{\rm iso}(xx_1)$ and comparing with
equation~(\ref{eq:annemi}) we obtain an extra check of
the consistency of our pair production and annihilation routines.

\subsection{Double Compton emissivity}

The angle-averaged double Compton spectral emissivity, 
$\ep_{\rm DC} (x,\Theta)\rmd x\;\ergs \cm^{-3} \sec^{-1} \sr^{-1}$, in an energy
interval $\rmd x$ is given by the expression (see, e.g., Svensson 1984): 
\be 
\epsilon_{\rm DC}(x,\Theta)=(n_{+}+n_{-})  x  
\frac{e^{-x/\Theta}}{2K_2(1/\Theta)}\int_0^{\infty} x_1^{-3} J(x_1) \rmd x_1 
\int_0^\infty \omega \frac{\rmd \sigma_{\rm DC}}{\rmd x}(x,\omega) 
 \exp\left( -\frac{\omega/x_1+x_1/\omega}{2\Theta}\right) \rmd \omega. 
\ee 
Here $J(x_1)$ is the mean intensity of the interacting photons. 
The differential cross section 
for the double Compton process is given in Svensson (1984,  eqs.~[A5], [A6]).
In order to account for the high energy cutoff at $x>\Theta$ 
 we introduced an {\it ad hoc}  exponential factor, $e^{-x/\Theta}$. 

\subsection{Bremsstrahlung emissivity}

The  emissivities due to relativistic electron-electron and positron-positron  
  thermal bremsstrahlung, $\ep_{\pm\pm} (x,\Theta) \rmd x\;
\ergs \cm^{-3} \sec^{-1} \sr^{-1}$, in an energy interval $\rmd x$ are given 
by the expression: 
\be 
\epsilon_{\pm\pm}(x,\Theta)=n_{\pm}n_{\pm} \sigmat\alpha_f \me c^3
 e^{-x/\Theta} \Theta^{-1/2} \frac{2^{3/2}}{3\pi} g_{\pm\pm}(x,\Theta) . 
\ee 
A similar expression holds for the electron-positron emissivity 
$\epsilon_{+-}(x,\Theta)$. 
We used the approximations by Skibo et al. (1995)  for the Gaunt factors
$g_{\pm\pm}(x,\Theta)$ and $g_{+-}(x,\Theta)$. Note that the term of unity 
in equations~(A7), (A9), and (A13)   in Skibo et al. (1995) should be deleted. 
  
\subsection{Numerical Integration} 

 All azimuthal integrations are made  using an 11-point Simpson quadrature. 
Furthermore, we apply 3-point Gaussian quadrature 
to calculate integrals over zenith angles for each hemisphere.  
The integration over frequencies is performed
using rectangular quadrature  on a logarithmic frequency scale, 
$\rmd x/x=\rmd \ln x$,   with   bin width $\sim 0.1$.  
The integrals over optical depth are calculated using rectangular quadrature.  
The number of points, $N_{\tau}$, is dependent on the geometry and the 
optical depth. We typically used $N_{\tau}=6$ for slabs, and $N_{\tau}=11-21$ 
for  active regions. 

\clearpage

\begin{deluxetable}{l l l r p{7mm} l l r}
\footnotesize
\tablecaption{Comparison of results obtained using the ISM and NLMC codes.}
\startdata
& & & & & & & \\ 
\tableline
&\multicolumn{3}{c}{SLAB} & \hspace{7mm} &\multicolumn{3}{c}{HEMISPHERE}\\
\tableline
Method & $\Theta$ & $\taut$ & $\ldiss$ & & $\Theta$ & $\taut$ & $\ldiss$ \\ 
\tableline 
ISM      & 0.19 & 0.29 & 260  &  & 0.24 & 0.70 & 560 \\
NLMC     & 0.19 & 0.29 & 300  &  & 0.24 & 0.70 & 700 \\
ISOSCAT1 & 0.19 & 0.29 & 1830 &  & 0.24 & 0.70 & 2730 \\
ISOSCAT2 & 0.19 & 0.29 & 390  &  & 0.24 & 0.70 & 740 \\
ISORAD   & 0.19 & 0.29 & 250  &  & 0.24 & 0.70 & 470 \\
ISOSF    & 0.19 & 0.30 & 340  &  & 0.24 & 0.76 & 640 \\
\noalign{\vskip 2mm}
ISM      & 0.29 & 0.17 & 19 &  & 0.49 & 0.29 & 27 \\
NLMC     & 0.29 & 0.16 & 20 &  & 0.49 & 0.28 & 40 \\
ISOSCAT1 & 0.29 & 0.17 & 86 &  & 0.49 & 0.29 & 70 \\
ISOSCAT2 & 0.29 & 0.17 & 25 &  & 0.49 & 0.29 & 32 \\
ISORAD   & 0.29 & 0.17 & 17 &  & 0.49 & 0.29 & 20 \\
ISOSF    & 0.29 & 0.17 & 23 &  & 0.49 & 0.30 & 31 \\
\noalign{\vskip 2mm}
ISM      & 0.82 & 0.036 & 0.24 &  & 1.20 & 0.073 & 1.5 \\
NLMC     & 0.82 & 0.033 & 0.20 &  & 1.20 & 0.070 & 3.0 \\
ISOSCAT1 & 0.82 & 0.037 & 0.47 &  & 1.20 & 0.076 & 2.4 \\
ISOSCAT2 & 0.82 & 0.036 & 0.27 &  & 1.20 & 0.073 & 1.6 \\
ISORAD   & 0.82 & 0.036 & 0.16 &  & 1.20 & 0.073 & 0.8 \\
ISOSF    & 0.82 & 0.036 & 0.27 &  & 1.20 & 0.075 & 1.6 

\enddata

\tablenotetext{}{
Input parameter: $\Theta \equiv k\Te/ \me  c^2$, the dimensionless pair 
temperature of the  corona (volume averaged in the NLMC case);
Output parameters: $\taut$, the (averaged) Thomson optical depth;
$\ldiss \equiv ( \Ldiss /H ) (\sigmat /  \me c^3)$, the local 
dissipation compactness of the coronal slab or hemisphere.
In all cases, $d=0$ and $\Tbb$ = 5 eV. 
ISM and NLMC represent results using the ISM and NLMC codes, respectively.
ISOSCAT1 and ISOSCAT2 represent ISM results using isotropic scattering
in the electron rest frame (Appendix A2).
ISORAD represent ISM results when  
the radiation field is made isotropic before solving the pair balance. 
ISOSF represent ISM results when the source functions, $S_{k\geq 2}$, are 
assumed to be isotropic and homogeneous (see \S~\ref{sect:isosf}). 
} 

\end{deluxetable}

\clearpage
 
\begin{table*}[h]
\begin{center}
\begin{tabular}{l l l l l l l} \hline
 Code &$\Theta$ & $\taut $ & $\ldiss$ &  & $\a218$ & \\ 
\tableline
ISM & &  & &  $\mu=0.113$ & $\mu=0.5$ & $\mu=0.887$ \\ 
HM93 & &  & &  $\mu=0.15$ & $\mu=0.55$ & $\mu=0.95$ \\ 
\tableline
ISM & 0.26 & 0.19 & 34 & 1.15 & 1.11 & 1.08 \\
HM93 & 0.26 & 0.20 & 22 & 1.23 & 1.14 & 1.10 \\
\noalign{\vskip 2mm}
ISM & 0.67 & 0.05 & 0.49 & 0.89 & 0.77 & 0.69 \\
HM93 & 0.67 & 0.05 & 2   & 0.87 & 0.73 & 0.65 \\
\noalign{\vskip 2mm}
ISM & 1.75 & 0.01 & 0.02 & 1.37 & 1.16 & 0.85 \\
HM93 & 1.75 & 0.01 & 0.1 & 1.23 & 0.95 & 0.53 \\ 
\tableline
\end{tabular}
\end{center}

\caption{Comparison of ISM results for coronal slabs with those of HM93}

\tablecomments{Input parameter:
$\Theta \equiv k\Te / \me c^2$, the dimensionless pair 
temperature of the  coronal slab;\\
Output parameters: 
$\taut$, the vertical Thomson optical depth of the coronal slab;\\
$\ldiss \equiv ( \Ldiss /H ) (\sigmat /  \me c^3)$, the dissipation compactness
of the coronal slab;\\
$\a218$, the least square fitted 2-18 keV intensity slope for
three specified cosine angles.\\
In all cases, $d=0$ and $\Tbb$ = 5 eV. 
See text for determination of $\ldiss$ from HM93.
}

\end{table*}

\clearpage

\end{document}